\documentclass[superscriptaddress,notitlepage,preprint,nofootinbib,amsmath,amssymb,floatfix]{revtex4-1}
\usepackage[caption=false]{subfig}
\usepackage[version=4]{mhchem}
\usepackage[english]{babel}
\usepackage{amsmath}

\usepackage{braket}
\usepackage{graphicx}
\usepackage{siunitx}
\usepackage{setspace}
\usepackage{float}
\usepackage{multirow}
\usepackage{wrapfig}
\usepackage{csvsimple}
\usepackage[toc,page]{appendix}

\usepackage{titlesec}
\usepackage{hhline}
\usepackage{longtable}
\usepackage{booktabs}
\usepackage[normalem]{ulem}
\usepackage{letltxmacro}
\usepackage{ebgaramond-maths}
\usepackage{natbib}
\usepackage[section]{placeins}
\usepackage{xcolor}
\LetLtxMacro{\ORIGselectlanguage}{\selectlanguage}
\makeatletter
\DeclareRobustCommand{\selectlanguage}[1]{%
	\@ifundefined{alias@\string#1}
	{\ORIGselectlanguage{#1}}
	{\begingroup\edef\x{\endgroup
			\noexpand\ORIGselectlanguage{\@nameuse{alias@#1}}}\x}%
}
\newcommand{\definelanguagealias}[2]{%
	\@namedef{alias@#1}{#2}%
}
\makeatother
\definelanguagealias{en}{english}
\definelanguagealias{EN}{english}

\newcommand{\aetfigsize}{0.8\columnwidth}
\AtBeginDocument{
\heavyrulewidth=.08em
\lightrulewidth=.05em
\cmidrulewidth=.03em
\belowrulesep=.65ex
\belowbottomsep=0pt
\aboverulesep=.4ex
\abovetopsep=0pt
\cmidrulesep=\doublerulesep
\cmidrulekern=.5em
\defaultaddspace=.5em
}
\begin{document}
\title{Ordering of Trotterization: Impact on Errors in Quantum Simulation of Electronic Structure}
\author{Andrew Tranter}
\affiliation{Department of Physics and Astronomy, Tufts University, Medford, MA 02155, USA}
\author{Peter J. Love}
\affiliation{Department of Physics and Astronomy, Tufts University, Medford, MA 02155, USA}
\author{Florian Mintert}
\affiliation{Department of Physics, Imperial College London, London SW7 2AZ, UK}
\author{Nathan Wiebe}
\affiliation{Department of Physics, University of Washington, Seattle, WA 98105, USA}
\affiliation{Pacific Northwest National Laboratory, Richland, WA 98382, USA}
\author{Peter V. Coveney}
\affiliation{Centre for Computational Science, University College London, London WC1H 0AJ, UK}
\date{\today}
\begin{abstract}
Trotter--Suzuki decompositions are frequently used in the quantum simulation of quantum chemistry.  They transform the evolution operator into a form implementable on a quantum device, while incurring an error---the Trotter error.  The Trotter error can be made arbitrarily small by increasing the Trotter number.  However, this increases the length of the quantum circuits required, which may be impractical.  It is therefore desirable to find methods of reducing the Trotter error through alternate means.
	
	The Trotter error is dependent on the order in which individual term unitaries are applied.  Due to the factorial growth in the number of possible orderings with respect to the number of terms, finding an optimal strategy for ordering Trotter sequences is difficult.  In this paper, we propose three ordering strategies, and assess their impact on the Trotter error incurred.
	
	Initially, we exhaustively examine the possible orderings for molecular hydrogen in a STO-3G basis.  We demonstrate how the optimal ordering scheme depends on the compatibility graph of the Hamiltonian, and show how it varies with increasing bond length.  We then use 44 molecular Hamiltonians to evaluate two strategies based on coloring their incompatibility graphs, while considering the properties of the obtained colorings.  We find that the Trotter error for most systems involving heavy atoms, using a reference magnitude ordering, is less than $1$ kcal/mol.  Relative to this, the difference between ordering schemes can be substantial, being approximately on the order of millihartrees.  The coloring-based ordering schemes are reasonably promising---particularly for systems involving heavy atoms---however further work is required to increase dependence on the magnitude of terms.  Finally, we consider ordering strategies based on the norm of the Trotter error operator, including an iterative method for generating the new error operator terms added upon insertion of a term into an ordered Hamiltonian. 
\end{abstract}
\maketitle
\section{Introduction}
Computational chemistry is the use of computers to answer outstanding questions in chemistry.  Various algorithms have been developed to calculate the properties of molecules and reactions.  The predictions of these methods are used in many fields.  For example, they are frequently used to aid development of synthetic processes~\cite{deglmannApplicationQuantumCalculations2015} and target searches in drug discovery~\cite{kortagereRoleComputationalMethods2012}.  Similarly, they can be used to characterize molecular configurations which are difficult to study experimentally, such as transition states of chemical reactions.
	
 The initial step of most computational approaches to chemistry involves some form of electronic structure theory calculation: the determination of molecular electronic wavefunctions and their corresponding energies~\cite{szaboModernQuantumChemistry1996,helgakerMolecularElectronicstructureTheory2000}.  Many methods for the solution of this problem have been established.
	
The most conceptually simple approach within this category is the full configuration interaction~(FCI) approach.  Here, a finite basis set of spin-orbitals is used to describe the Hilbert space of the electronic wavefunction.  Typically, this is initially a localized basis of atomic orbitals.  Molecular orbitals are found using an algorithm such as the Hartree--Fock method, with the Hamiltonian in a basis of Slater determinants formed from these molecular orbitals giving the configuration interaction (CI) matrix.  The eigenstates and eigenvalues of the CI matrix then give electronic eigenstates and their corresponding energies.  Such a technique is numerically exact, to within the limitations of the basis set and the assumption that relativistic effects are negligible.  However, the computational expense of this technique scales factorially with the number of basis functions used~\cite{szaboModernQuantumChemistry1996}.  This limits the application to extremely small molecules~\cite{ganLowestEnergyStates2006}, and thus is typically used as a benchmark for other methods.
	
Approximate methods such as coupled cluster theory~\cite{popleElectronCorrelationTheories1978} or M\o ller--Plesset perturbation theory~\cite{mollerNoteApproximationTreatment1934} are often used to obtain results with practical computational resources; however, exact methods are required for benchmarking these.  Additionally, for any approximation applied, a system can be found wherein such an approximation breaks down.  These facts reinforce the necessity for computationally feasible numerically exact methods.
	
	It has been established that a scalable quantum computer would be capable of providing full configuration interaction level electronic structure results in polynomial time~\cite{aspuru-guzikSimulatedQuantumComputation2005,kassalSimulatingChemistryUsing2011,caoQuantumChemistryAge2018}.  As quantum chemistry is expected to be a key application of developing quantum devices~\cite{olsonQuantumInformationComputation2017}, there has been a great deal of theoretical development on the algorithms that would be used to perform quantum chemical calculations on a scalable quantum computer.  Algorithms to describe various chemical processes have been developed, including energy spectra~\cite{aspuru-guzikSimulatedQuantumComputation2005}, reaction rates~\cite{lidarCalculatingThermalRate1999,kassalQuantumAlgorithmMolecular2009,reiherElucidatingReactionMechanisms2017} and reaction dynamics~\cite{kassalPolynomialtimeQuantumAlgorithm2008}.  Experimental demonstrations have been shown on a variety of quantum computing architectures, including photonic~\cite{lanyonQuantumChemistryQuantum2010}, nuclear magnetic resonance~\cite{duNMRImplementationMolecular2010},  superconducting~\cite{omalleyScalableQuantumSimulation2016,kandalaHardwareefficientVariationalQuantum2017} and trapped-ion~\cite{hempelQuantumChemistryCalculations2018,namGroundstateEnergyEstimation2019} systems.
	
	The variational quantum eigensolver (VQE) algorithm~\cite{peruzzoVariationalEigenvalueSolver2014,mccleanTheoryVariationalHybrid2016} is a hybrid quantum-classical scheme, where a classical variational approach is taken with a quantum device used to determine accurate expectation values.  This approach has obtained much attention in the recent literature, although other hybrid quantum-classical schemes for the simulation of quantum systems  \cite{smartQuantumclassicalHybridAlgorithm2019,weiFullQuantumEigensolver2019} and other purposes~\cite{farhiQuantumApproximateOptimization2014,iliyasuHybridQuantumClassicalProtocol2014,zhuTrainingQuantumCircuits2019} have been reported. Although repeated Ansatz preparation and the need for multiple variational steps increases the overall asymptotic cost of these algorithms, they do not require coherency to be maintained throughout the entire circuit, instead requiring many short coherent evolutions.  As such, while not providing full configuration interaction level accuracy, these methods can efficiently provide results that are more accurate than classical equivalents.  These methods allowed for the simulation of beryllium hydride in a minimal basis in 2017, albeit with low accuracy~\cite{kandalaHardwareefficientVariationalQuantum2017}.
		
	Although recent experimental approaches have focused on the variational quantum eigensolver, open questions remain in the study of approaches based on a phase estimation algorithm~\cite{kitaevQuantumMeasurementsAbelian1995,lloydUniversalQuantumSimulators1996,abramsSimulationManyBodyFermi1997, aspuru-guzikSimulatedQuantumComputation2005}. This requires coherency to be maintained for a time that scales exponentially with the desired precision.  While this is not asymptotically prohibitive due to the required precision being fixed, it is markedly less practical for implementation on noisy devices.  However, this does result in eigenvalues which are numerically exact.
	
	Trotterization is the use of Trotter--Suzuki formulae to simulate evolution under a given Hamiltonian which comprises a sum of many terms, by sequentially simulating the evolution under each term.  In both phase estimation and VQE, Trotterization of the qubit Hamiltonian is a basic tool.  It can be used for the time evolution required by phase estimation and for Ansatz preparation in VQE.  In doing this, a degree of error---the \emph{Trotter error}---is introduced.  This error can be made arbitrarily small by increasing the Trotter number, i.e. the number of times the terms are iterated through.  However, this increases the quantum computational cost of the procedure.  As it is likely that quantum computational resources will be highly limited in the foreseeable future, it is useful to find methods to reduce this overhead in order to enable the simulation of larger systems.  This is particularly relevant when considering phase estimation approaches, as Trotter error in these translates to the only source of algorithmic error in the simulation, whereas the impact of this error in a variational scheme is minimized through optimization of Ansatz parameters~\cite{romeroStrategiesQuantumComputing2018}.
	
	It has been shown~\cite{hastingsImprovingQuantumAlgorithms2015,tranterComparisonBravyiKitaev2018} that the error incurred in the use of a Trotter--Suzuki approximation is dependent on the order in which individual terms are simulated.  However, the use of a Trotter ordering scheme is not determined by the Trotter error effects alone.  Differing Trotter ordering can have an effect on the length of the circuit required to simulate one Trotter step, as identical gates between terms can be canceled~\cite{hastingsImprovingQuantumAlgorithms2015}, with the degree of cancellation being dependent on the ordering of terms.  Several ordering schemes have been reported to this end  \cite{hastingsImprovingQuantumAlgorithms2015,poulinTrotterStepSize2015,babbushChemicalBasisTrotterSuzuki2015,tranterComparisonBravyiKitaev2018}.    Nonetheless, it is possible that this effect could be outweighed by the minimization of Trotter error, and the consequent reduction in the overall number of Trotter steps---especially where the Trotter number necessary for chemically accurate predictions is low, or the ordering impact on Trotter error is high.  In this paper, we thus focus on attempts to reduce the single-Trotter-step Trotter error, instead of considering other effects that impact circuit length.  To this end, we report and characterize two new ordering strategies.
	
 	It is possible to derive an analytic expression for the \emph {Trotter error operator}~\cite{poulinTrotterStepSize2015}, of which the expectation value in the ground state yields the Trotter error.  As finding the exact ground state is exponentially hard, finding the exact Trotter error with this approach is similarly difficult.  However, the norm of the Trotter error operator can be determined without diagonalizing the full Hamiltonian.  This serves as an upper bound to the true Trotter error, although this bound can be extremely loose~\cite{babbushChemicalBasisTrotterSuzuki2015}.  It has been observed~\cite{babbushChemicalBasisTrotterSuzuki2015} that,  despite the looseness of this bound, it does replicate some qualitative trends of the true Trotter error.  As such, it is possible that using this information to guide an ordering scheme could result in an effective strategy.  In Section~\ref{sec:errorOperator}, we report and assess an ordering scheme in this vein.  While the Trotter error operator can be computed in polynomial time, it remains computationally intensive.  We therefore derive a Trotter term insertion error operator, which yields the terms that are added to the Trotter error operator upon insertion of a new term into the Hamiltonian.  This allows direct and efficient comparison of the impact upon the Trotter error operator, of inserting a term into the Hamiltonian in different positions.  This is of use in the greedy ordering scheme discussed in Section~\ref{sec:errorOperator}.
 	
	We begin by providing a brief overview of the theory underpinning the canonical methods of the quantum simulation of electronic structure theory, and particularly the theory of Trotterization~\cite{aspuru-guzikSimulatedQuantumComputation2005}.  We then consider three approaches to the development of Trotter ordering schemes.  Firstly, in Section~\ref{sec:hydrogen}, we discuss the hydrogen molecule in a minimal basis as a simple case study.  Although this system has been extensively studied, including the consideration of all possible Trotter orderings in an experimental context~\cite{omalleyScalableQuantumSimulation2016}, an examination of the distribution of Trotter errors across varying Trotter orderings is yet unreported in the literature.  We consider this here, along with a discussion of the geometry dependence of the optimal Trotter ordering.  Secondly, in Section~\ref{sec:generalised}, we propose two ordering schemes based on subdividing the molecular Hamiltonian into mutually commuting subsets of terms, and report their performance across a dataset of 44 systems.  Finally, in Section~\ref{sec:errorOperator}, we propose and assess a final ordering scheme based on minimizing the norm of the Trotter error operator.

	\section{Trotterization---Theoretical Background}
	The electronic Hamiltonian in the second quantized formalism is given by:
	
		\begin{equation}
		\hat{H} = \sum_{i,j}{h_{ij}a^\dagger_i a_j}+\frac{1}{2} \sum_{i,j,k,l}{h_{ijkl} a^\dagger_i a^\dagger_j a_k a_l}
		\label{eq:elecham}
		\end{equation}
		
	where $h_{ij}$ and $h_{ijkl}$ are Coulombic overlap and exchange integrals determined by the basis set chosen~\cite {helgakerMolecularElectronicstructureTheory2000,szaboModernQuantumChemistry1996}.  The overall goal of the simulation process is to determine the eigenstates and corresponding eigenvalues of this Hamiltonian.  Letting the number of spin-orbitals included in the basis set be $N$, there are $O(N^4)$ terms in the Hamiltonian~in Equation \eqref{eq:elecham}.  In practice, molecular symmetries and orbital localization result in many of these terms having a zero or negligible coefficient.  Regardless, the determination of the $h_{ij}$ and $h_{iijkl}$ integrals is classically efficient, and thus these weighting constants can be considered as input data when performing the calculation on either a classical or quantum~device.	

	Classically, the matrix elements of the electronic Hamiltonian in a basis of Slater determinants may be obtained.  The eigenvalues of the full configuration interaction matrix can then be determined, for instance by direct diagonalization.  However, the dimension of the Fock space which $\hat{H}$ acts upon grows exponentially with $N$.  This can be reduced by excluding the subspace with the incorrect number of electrons; however, this still results in a growth of $\binom{N}{n}$, where $n$ is the number of electrons.  As such, it is intractable to perform this process for more than a handful of tens of spin-orbitals.  Conversely, $N$ qubits span the entire Fock space that $\hat{H}$ acts upon---a quantum computer thus circumvents the exponential cost of the simulation procedure.

	Performing a similar procedure on a quantum device differs from a classical full configuration interaction calculation.  Having calculated the $h_{ij}$ and $h_{ijkl}$ integrals, a mapping scheme to transform the creation and annihilation operators of Equation    \eqref{eq:elecham}, along with the electronic states they act upon, to operations upon and states of qubits must be found.  This is typically performed through the Jordan--Wigner transformation, the Bravyi--Kitaev transformation, or other constructions~\cite{jordanUberPaulischeAquivalenzverbot1928,bravyiFermionicQuantumComputation2002,seeley_bravyi-kitaev_2012-5,tranterBravyiKitaevTransformationProperties2015,tranterComparisonBravyiKitaev2018,setiaBravyiKitaevSuperfastSimulation2017} . In this paper, we primarily utilize the Jordan--Wigner transformation, although we present some results using the Bravyi--Kitaev mapping in Section~\ref{sec:generalised}.  The result of performing this mapping procedure is the generation of a qubit Hamiltonian, which consists of a sum of weighted strings of qubit Pauli operators.

From here, two approaches to finding molecular eigenstates and eigenvalues are common.  In quantum phase estimation~\cite{kitaevQuantumMeasurementsAbelian1995,nielsenQuantumComputationQuantum2010}, a circuit corresponding to the the evolution operator $U = \exp\left(-it\hat{H}\hbar\right)$ of the qubit Hamiltonian is repeatedly applied, with a second register of qubits used to store a binary expansion of the true eigenvalue.  In a variational quantum eigensolver, a parameterized Ansatz state which is classically hard to store is generated.  The expectation value of the qubit Hamiltonian with this state is measured, and a classical optimizer used to vary the Ansatz parameters until the expectation value is variationally minimized~\cite{peruzzoVariationalEigenvalueSolver2014,mccleanTheoryVariationalHybrid2016}.  Typically, the unitary coupled cluster Ansatz~\cite{romeroStrategiesQuantumComputing2018}---or a form derived from it~\cite{leeGeneralizedUnitaryCoupled2019}---is used.  Here, an exponentiated form of the unitary cluster operator is applied to a reference state.

In both cases, a circuit that implements an exponentiated sum of Pauli operators must be found.  In general, such a circuit is difficult to find.  However, standard circuits exist to simulate individual Pauli strings---$U_i = \exp\left(-it_i\hat{H}_i\hbar\right)$.  As such, we invoke a Trotter--Suzuki approximation to break the overall evolution operator into a product of individual terms.

A first-order Trotter--Suzuki approximation is given by~\cite{hatanoFindingExponentialProduct2005}:

\begin{equation}
e^{-i\skew{3.8}\hat{H}t} \approx \left( \prod^m_{k=1}{e^{-i\skew{3.8}\hat{H}_kt/N_{T}}}     \right)^{N_{T}}\,.
\end{equation}

where the Hamiltonian $\skew{3.8}\hat{H}$ is a sum of $m$ terms $\skew{3.8}\hat{H}_k$.  Intuitively, this states that the Hamiltonian can be approximated by rapidly switching between each term, across the desired evolution time.  The exponential of a single Hamiltonian term can be described easily.  Increasing $N_{T}$, the \emph{Trotter number}---the number of Trotter steps---deterministically reduces the error incurred by this approximation, but with commensurate increase in circuit length.

A second-order Trotter--Suzuki approximant 

\begin{equation}
e^{-i\skew{3.8}\hat{H}t} \approx \left( \prod^m_{k=1}{e^{-i\skew{3.8}\hat{H}_kt/2N_{T}}}       \prod^1_{k=m}{e^{-i\skew{3.8}\hat{H}_kt/2N_{T}}}     \right)^{N_{T}}
\label{trotter1}
\end{equation}

can be obtained by symmetrizing the first-order approximant.  Higher-order approximants can be obtained recursively~\cite{suzukiGeneralizedTrotterFormula1976}.  Increasing the approximation order will yield reductions in error.  However, the overall length of the quantum circuit required to simulate higher-order Trotter approximants increases exponentially with the order used.  Choosing an appropriate strategy for simulation is then a three-way trade-off  among  the order of Trotter--Suzuki approximant, the number of time steps, and the need to minimize the overall circuit length.  For consistency with other work, we mostly restrict ourselves here to use of a second-order Trotter approximant (Equation    \eqref{trotter1}), although we additionally use a first-order Trotter approximant in Section~\ref{sec:hydrogen}.

   To bound the error incurred in the use of this approximation, \citet{poulinTrotterStepSize2015} introduced the Trotter \emph{error operator}, which gives the Trotter error for a given Trotterized Hamiltonian. For a second-order Trotter--Suzuki approximant, the expectation value of this with an eigenstate is given by~\cite{babbushChemicalBasisTrotterSuzuki2015}:

\begin{equation}
\Delta E_i = -\frac{\Delta^2_t}{12}\sum_{\alpha \le \beta}{\sum_{\beta}{\sum_{\gamma < \beta}{\bra{\psi_i}  \left[  \skew{3.8}\hat{H}_\alpha \left(  1- \frac{\delta_{\alpha,\beta}}{2}            \right)  ,\left[   \skew{3.8}\hat{H}_\beta,\skew{3.8}\hat{H}_\gamma \right]           \right] \ket{\psi_i}}}}\,,
\label{errorOperator}
\end{equation}

where $	\Delta E_i$ is the expected error (ignoring higher-order terms), $\Delta^2_t$ is the Trotter step size, $\ket{\psi_i}$ is the ground state , $\delta$ is the Kronecker delta function and $\skew{3.8}\hat{H}_i$ are Hamiltonian terms~\cite{babbushChemicalBasisTrotterSuzuki2015}.

Examining this, it is evident that two factors primarily affect the degree of error introduced by the use of the Trotter--Suzuki approximation.  Firstly, the error will decrease with the number of time steps used.  Additionally, as the sums in Equation    \eqref{errorOperator} are ordered---being sums over conditional indices, rather than the whole range---the order in which terms are applied will impact the error incurred.  Indeed, previous work has suggested that the Trotter ordering can dramatically impact the number of Trotter steps required for constant precision~\cite{seeley_bravyi-kitaev_2012-5,hastingsImprovingQuantumAlgorithms2015}.  It is thus desirable to find an ordering strategy which minimizes this error, motivating this paper.

\color{black}

In this section, we have discussed the underlying theory of Trotterization and the impact of Trotter ordering schemes.  For most molecular systems, the space of possible orderings is sufficiently vast as to inhibit direct statistical analysis.  We proceed to consider a small test case---molecular hydrogen in a minimal basis---both to justify the generalized ordering schemes presented in Section~\ref{sec:generalised} and to study the role of molecular geometry in a small example.

\section{Molecular Hydrogen}
\label{sec:hydrogen}
Molecular hydrogen in an STO-3G basis is the smallest and simplest chemically interesting electronic structure problem---barring its cationic form, which lacks two-electron interactions.  Due to its simplicity, this example has been widely used both experimentally and theoretically~\cite{aspuru-guzikSimulatedQuantumComputation2005,seeley_bravyi-kitaev_2012-5,setiaBravyiKitaevSuperfastSimulation2017,lanyonQuantumChemistryQuantum2010,duNMRImplementationMolecular2010,omalleyScalableQuantumSimulation2016,hempelQuantumChemistryCalculations2018}.  It was recently shown that this example is in a sense not quantum mechanical, as it lacks measurement contextuality~\cite{kirbyContextualityTestNonclassicality2019}; however, the simplicity of the system makes it a good candidate for initial studies.

The qubit Hamiltonian for the hydrogen molecule in an STO-3G basis, using a Jordan--Wigner transformation, with a bond length of~\SI{0.7414}{\angstrom}, is given by:
	
		\begin{align}\skew{3.8}\hat{H_q} = &-0.81262 I + 0.17120 \sigma^z_0 + 0.17120 \sigma^z_1 - 0.22279\sigma^z_2 \nonumber \\ 
&- 0.22279\sigma^z_3 + 0.16862\sigma^z_1\sigma^z_0 + 0.12054\sigma^z_2\sigma^z_0 + 0.16587\sigma^z_3\sigma^z_0 \nonumber \\ 
&+ 0.16587\sigma^z_2\sigma^z_1 + 0.12054\sigma^z_3\sigma^z_1 + 0.17435 \sigma^z_3\sigma^z_2 \nonumber \\
&- 0.04532\sigma^y_3\sigma^y_2\sigma^x_1\sigma^x_0 + 0.04532\sigma^x_3\sigma^y_2\sigma^y_1\sigma^x_0 \nonumber \\
&+ 0.04532\sigma^y_3\sigma^x_2\sigma^x_1\sigma^y_0 - 0.04532\sigma^x_3\sigma^x_2\sigma^y_1\sigma^y_0\,.
\label{hyHam}
\end{align}

where the lower index of each Pauli operator corresponds to the index of the qubit it is applied to.  This consists of 15 terms.  Unfortunately, even a Hamiltonian of this small size provides on the order of $10^{12}$ possible orderings.  It is thus infeasible to consider all orderings for the full Hamiltonian, even for the hydrogen molecule in a minimal basis.  The second-order Trotter error operator (Equation    \eqref{errorOperator}) consists of a sum of many triple commutators between Hamiltonian terms.  As such, it is prudent to consider the commutativity of the terms in the Hamiltonian.  Inspecting Equation    \eqref{hyHam} shows that the terms with only two Pauli Z operators commute with all terms in the Hamiltonian.  Along with the identity term, these  form a \emph{totally commuting set}.  Because they commute with all terms, their exponentials commute with the exponentials of all other terms, and consequentially they can be freely moved around a Trotterized evolution operator.  As such, all terms in this set can be moved to the front of the operator, combined, and simulated as an entirely independent operator

\begin{equation}
e^{\frac{-i\skew{3.8}\hat{H}t}{\hbar}} \approx e^{\frac{-i\skew{3.8}\hat{H}_{C}}{\hbar}}T\left(\skew{3.8}\hat{H}_{A}\right)\,,
\end{equation}

where $\skew{3.8}\hat{H}_{C}$ is the sum of all totally commuting terms, and $T\left(\skew{3.8}\hat{H}_{A}\right)$ is a Trotterization of the other terms.  In other words, we only need to Trotterize the terms that do not commute with all terms.

This analysis has two principal advantages.  Firstly, as the totally commuting set does not require a Trotter--Suzuki approximation, these terms can be simulated in one time step.  This dramatically cuts down on the number of gates required for implementation on a quantum device.  It should be noted here that for larger systems the size of the totally commuting set rapidly drops to one term (the identity operator).  As such, this advantage is not scalable, and therefore this technique cannot be used directly in the analysis of larger systems.

More importantly for the purposes of this discussion, as we do not need to Trotterize the terms in the totally commuting set, the space of possible orderings is dramatically reduced to $40,320$.  As the simulation of each ordering takes less than 1 s on an average laptop computer, it is thus feasible to simply brute force search the entire space of orderings in order to study the distribution of errors. 

In our simulations, integral data  were generated using the Psi4~\cite{parrishPsi4OpenSourceElectronic2017} quantum chemistry package and OpenFermion~\cite{mccleanOpenFermionElectronicStructure2017}, with a standard Hartree--Fock basis used to express the molecular Hamiltonian.  Our Python code was used to generate both Jordan--Wigner and Bravyi--Kitaev Hamiltonians.  The exact ground state and its corresponding energy were then determined.  Similarly, the Trotter error for every possible ordering was determined, with the overall evolution time and the number of Trotter steps set to unity.

\begin{figure}[H]
	\centering
    {\includegraphics[width=\aetfigsize]{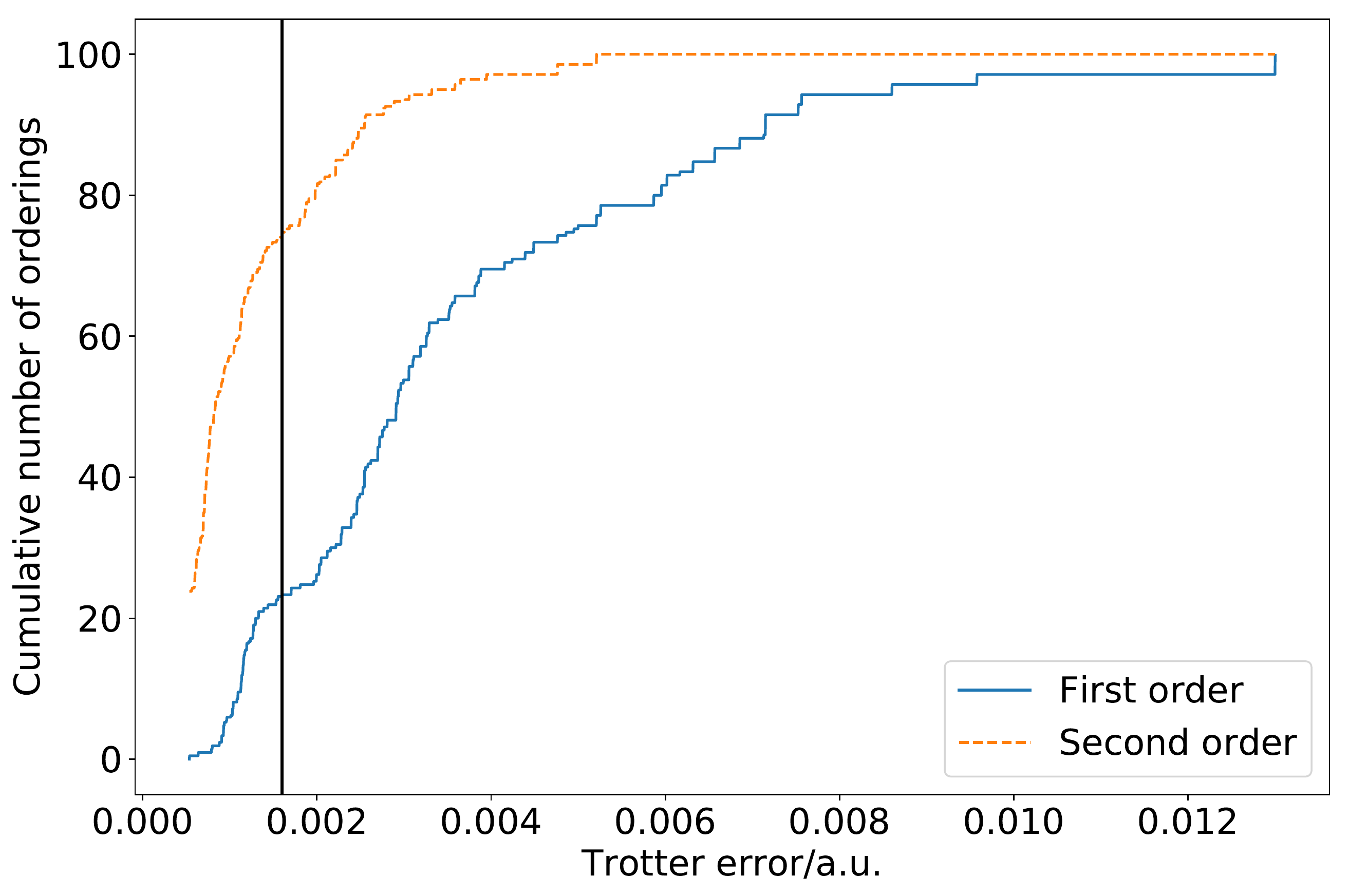}}	
    \caption[Cumulative density plot of hydrogen ordering errors]{Cumulative density plot of hydrogen ordering errors for one Trotter step, using a bond length of~\SI{0.7414}{\angstrom} and an STO-3G atomic basis.  The vertical line denotes an error of $1$ kcal/mol.  Approximately $20\%$ of the orderings achieve this error or lower for the first-order Trotter--Suzuki approximation.  Around $80\%$ of orderings achieve an error of 0.005 Hartree, approximately half that of the worst possible ordering.}
	\label{hydrogenOrderings}
\end{figure}

Figure~\ref{hydrogenOrderings} is a cumulative density plot showing the distribution of Trotterization errors which are obtained as a result of ordering variations.  Errors are given relative to the true eigenvalue of the Hamiltonian with totally commuting terms removed.  The same distribution is obtained regardless of whether the Jordan--Wigner or Bravyi--Kitaev Hamiltonian is simulated.  The distribution is heavily weighted towards the low error region.  This is promising, as it implies that a random choice of ordering is likely to introduce a small amount of ordering-dependent error.  There is, however, over an order of magnitude difference between the optimal and poorest orderings.  In this case, the difference is such that only three Trotter steps are required to reach an accuracy of $0.0001 $   a.u. for the optimal ordering, whereas seven Trotter steps are required for the worst ordering.  This implies that---if we extrapolate solely from the hydrogen molecule in a minimal basis---picking a bad ordering could result in a dramatically increased gate count in actual quantum simulations.  This is important despite the high likelihood of random orderings being accurate, as if a systematic approach to ordering is taken, it must be ensured that this does not result in bad orderings being selected.  It should be noted here, however, that the results of Section~\ref{sec:generalised}   show that this difference is lessened in systems involving heavier~atoms.

\begin{figure}[]
	\centering
    {\centering \includegraphics[width=0.6\textwidth]{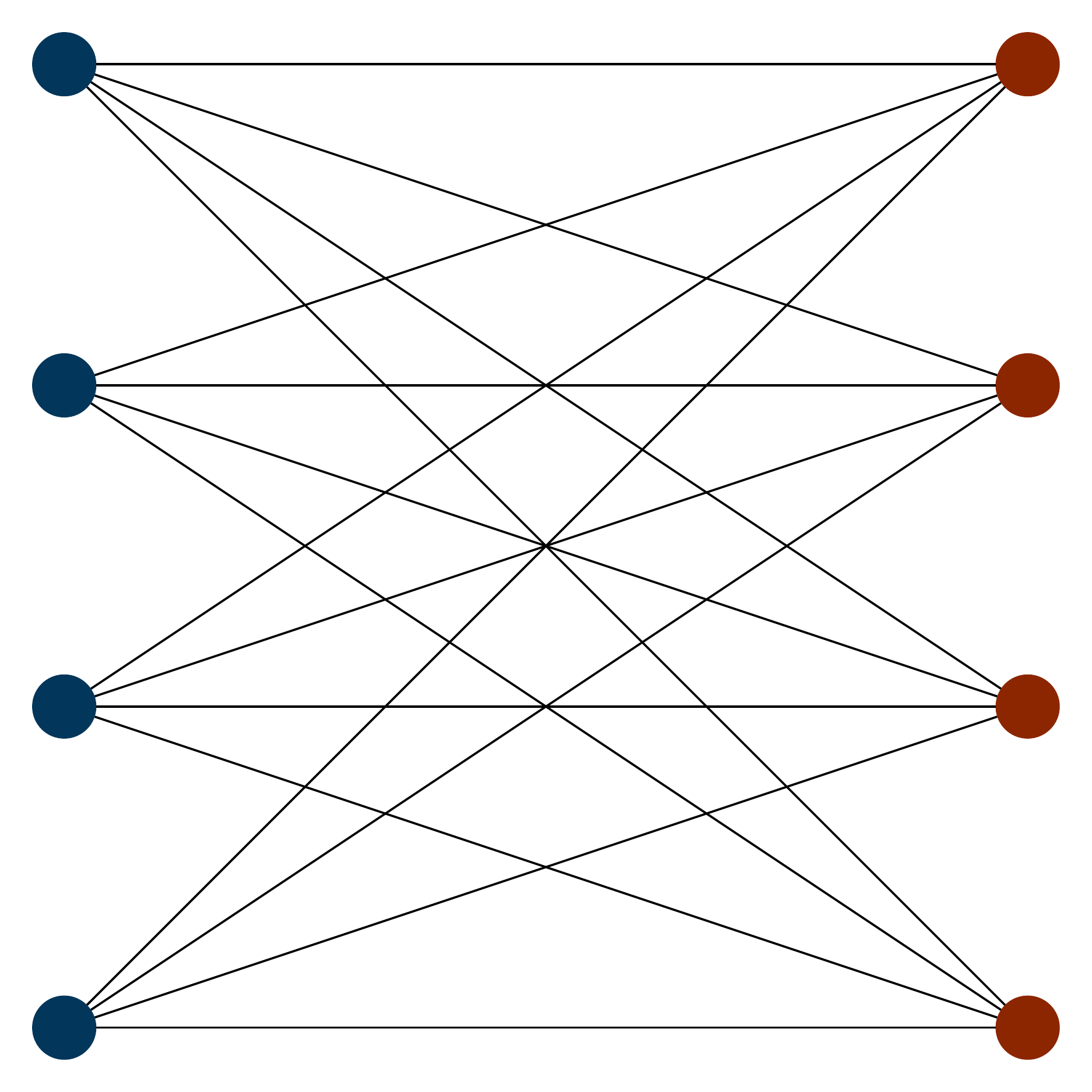}}
       \caption[Further representations of the commutativity graph of the hydrogen Hamiltonian]{The incompatibility graph of the Jordan--Wigner and Bravyi--Kitaev Hamiltonians, with the totally commuting set removed.  Nodes correspond to Hamiltonian terms, edges correspond to non-commutativity between terms.  Two independent sets are clearly revealed, with the XY-set colored blue and the Z-set colored red.}
           \label{fig:hyCommutatitivity2}
\end{figure}

We now consider the optimal ordering strategy for the hydrogen molecule, by re-examining the commutativity structure graph (the incompatibility graph) discussed above.  Figure~\ref{fig:hyCommutatitivity2} shows this with totally commuting terms removed. Here, it is apparent that the incompatibility graph is bipartite.  As such, the Hamiltonian can be subdivided into two sets of terms.  Within each set, all terms mutually commute.  No term in a given set commutes with any term in the other set.  One set consists of all of the terms with Pauli Z operators, whereas the other set consists of terms with Pauli X and Y operators.  We label the former of these sets the Z-set and the latter the XY-set.  Using the Jordan--Wigner transformation, the optimal ordering using a first-order Trotter expansion is given by
\begin{align*}
\skew{3.8}\hat{H}_1 =  
&{\color[rgb]{0.008,0.212,0.353}0.04532\sigma^x_3\sigma^y_2\sigma^y_1\sigma^x_0} 
{\color[rgb]{0.549,0.149,0} -0.22280\sigma^z_2} 
{\color[rgb]{0.008,0.212,0.353} -0.04532\sigma^y_3\sigma^y_2\sigma^x_1\sigma^x_0}\\
& {\color[rgb]{0.549,0.149,0} - 0.22280\sigma^z_3} 
{\color[rgb]{0.008,0.212,0.353} - 0.04532\sigma^x_3\sigma^x_2\sigma^y_1\sigma^y_0}
 {\color[rgb]{0.549,0.149,0} + 0.17120 \sigma^z_1}\\
& {\color[rgb]{0.008,0.212,0.353} + 0.04532\sigma^y_3\sigma^x_2\sigma^x_1\sigma^y_0 }
 {\color[rgb]{0.549,0.149,0} + 0.17120 \sigma^z_0}\,,
\end{align*}
where the leftmost term is simulated first, the second term is simulated second, and so on.  Here, the first, third, fifth and seventh terms are members of the XY-set, while the remaining terms are members of the Z-set.  Examining this, it is evident that the optimal ordering strategy is given by alternating between completely commuting sets in order of descending coefficient magnitude.  This strategy is optimal for both Jordan--Wigner and Bravyi--Kitaev Hamiltonians.  However, other systematic approaches to ordering strategies for this Hamiltonian produce differing results for the two mapping techniques.  This emphasizes the need for a systematic ordering scheme which is mapping agnostic.

\subsection{Geometry Dependence}

\begin{figure}[]
	\centering
    {\centering \includegraphics[width=0.6\textwidth]{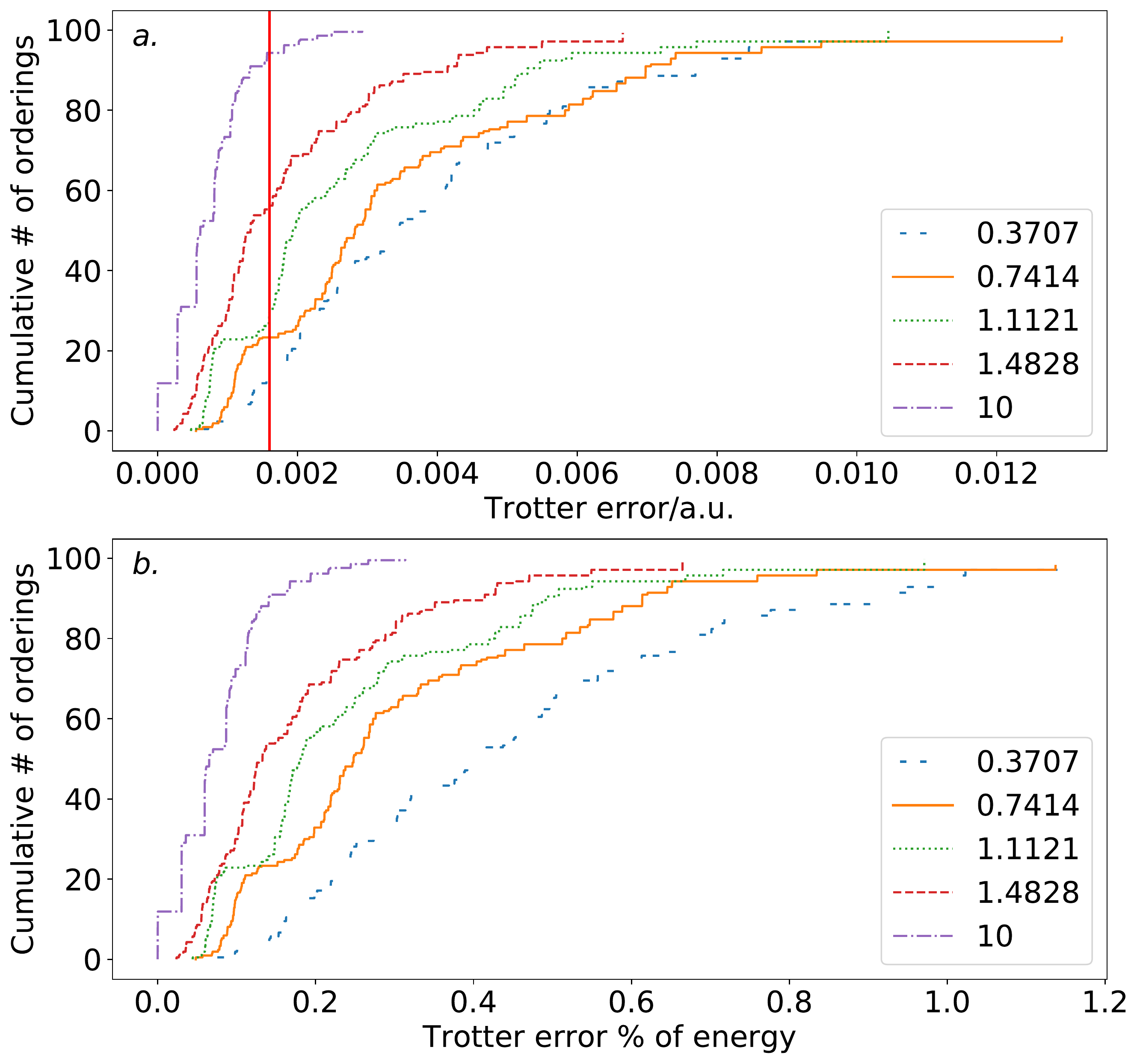}}
        \caption{Distribution of Trotter errors by ordering for varying bond lengths for \ce{H2} in a minimal basis.  a: versus absolute Trotter error.  As the bond length decreases, both the Trotter error and the dependence of the Trotter error on the ordering chosen increase.  b: versus the Trotter error as a percentage of the ground state energy.  The same trend as with the absolute Trotter error is observed, although the ordering dependence at extremely low bond length is accentuated.}
    \label{fig:hydrogenBondlengths}
\end{figure}

Molecular hydrogen allows for the specification of the entire molecular geometry with a single parameter; the bond length of the molecule.  This allows for the consideration of how the Trotter error varies with the molecular geometry.  Figure~\ref{fig:hydrogenBondlengths} demonstrates this.  To contrast the results above and for consistency with prior work, we here use a second-order Trotter expansion.  Increasing the bond length results in a substantially reduced Trotter error.  This is likely due to the increased locality of the electronic eigenstates, resulting in increased degeneracy, and thus in increased symmetry in the coefficients of the Z-set terms.  At asymptotic separation, the Z-set terms have equal coefficient.  This additional symmetry reduces the dependence of the Trotter error on ordering choice.  It also allows for increased cancellation in the terms of the Trotter error operator, resulting in a reduction in Trotter error overall at higher bond lengths.  

Table~\ref{tab:hydrogenBondLengths} shows how the Pauli Hamiltonian and optimal Trotter ordering varies with bond length.  We first observe that the optimal ordering has changed at equilibrium bond length, due to the use of the second-order Trotter--Suzuki formula.  The strategy of alternating between commuting sets remains.  However, within each set, the terms are ordered in descending magnitude order, likely due to the symmetrization of the second-order approximant.

For small bond length, we observe substantially increased difference in the coefficients of the Z-set terms, due to the increased relative stability of the bonding orbitals.  The strategy of alternating between Z-set and XY-set terms is optimal at all pre-asymptotic separations.  However, the particular choice of Z-set terms differs at higher bond lengths, due to the increasing similarity of the coefficients of the Z-set terms.  At asymptotic separation, a different ordering is preferable.  Here, as the Z-set terms are equal, the Trotter error can be reduced to zero by placing all XY-set terms at the start of the expansion.

\begin{table}[]
\centering
\caption{Optimal Trotter orderings for the Hydrogen molecule in a STO-3G basis, for varying bond length, using a second-order Trotter--Suzuki approximation and a Jordan--Wigner mapping.  Each ordering proceeds from top to bottom.  The ordering changes as bond length increases, although, prior to the asymptotic limit, all orderings are of the form of alternating between commuting sets.  At asymptotic separation, it is preferable to simulate sets in sequence, due to the symmetry of the coefficients.}
\label{tab:hydrogenBondLengths}
\begin{tabular}{ccccc}
\toprule
\multicolumn{5}{c}{\bf{ Bond length (\AA)}} \\ 
\midrule \multicolumn{1}{c}{\bf {0.3707}} & \multicolumn{1}{c}{\bf{0.7414}} & \multicolumn{1}{c}{\bf{1.1121}} & \multicolumn{1}{c}{\bf 1.4828} & \multicolumn{1}{c}{\bf 10.000} \\
\midrule
0.24197 $\sigma^z_1$ & 0.17120 $\sigma^z_1 $ & $-$0.10205 $\sigma^z_2$ & $-$0.03780 $\sigma^z_2$ & 0.09021 $\sigma^y_3 \sigma^x_2 \sigma^x_1 \sigma^y_0$  \\

0.04084 $\sigma^y_3 \sigma^x_2 \sigma^x_1 \sigma^y_0$ & 0.04532 $\sigma^y_3 \sigma^x_2 \sigma^x_1 \sigma^y_0$ & 0.05100 $\sigma^y_3 \sigma^x_2 \sigma^x_1 \sigma^y_0$ & 0.05711 $\sigma^y_3 \sigma^x_2 \sigma^x_1 \sigma^y_0$ & 0.09021 $\sigma^x_3 \sigma^y_2 \sigma^y_1 \sigma^x_0$  \\

 0.24197 $\sigma^z_0$ & 0.17120 $\sigma^z_0$ & $-$0.10205 $\sigma^z_3$ & $-$0.03780 $\sigma^z_3$ & $-$0.09021 $\sigma^x_3 \sigma^x_2 \sigma^y_1 \sigma^y_0$ \\

 $-$0.04084 $\sigma^y_3 \sigma^y_2 \sigma^x_1 \sigma^x_0$ & $-$0.04532 $\sigma^y_3 \sigma^y_2 \sigma^x_1 \sigma^x_0$ & 0.05100 $\sigma^x_3 \sigma^y_2 \sigma^y_1 \sigma^x_0$ & 0.05711 $\sigma^x_3 \sigma^y_2 \sigma^y_1 \sigma^x_0$& $-$0.09021 $\sigma^y_3 \sigma^y_2 \sigma^x_1 \sigma^x_0$ \\

$-$0.48079 $\sigma^z_3$ & $-$0.22279 $\sigma^z_3$ &  0.12533 $\sigma^z_0$ & 0.09462 $\sigma^z_0$& 0.03964 $\sigma^z_0$\\

$-$0.04084 $\sigma^x_3 \sigma^x_2 \sigma^y_1 \sigma^y_0$ & $-$0.04532 $\sigma^x_3 \sigma^x_2 \sigma^y_1 \sigma^y_0$ & $-$0.05100 $\sigma^y_3 \sigma^y_2 \sigma^x_1 \sigma^x_0$ & $-$0.05711 $\sigma^y_3 \sigma^y_2 \sigma^x_1 \sigma^x_0$ &0.03964 $\sigma^z_1$ \\

$-$0.48079 $\sigma^z_2$ &$-$0.22279 $\sigma^z_2$  & 0.12533 $\sigma^z_1 $ & 0.09462 $\sigma^z_1 $ & 0.03964 $\sigma^z_3$\\

 0.04084 $\sigma^x_3 \sigma^y_2 \sigma^y_1 \sigma^x_0$ & 0.04532 $\sigma^x_3 \sigma^y_2 \sigma^y_1 \sigma^x_0$ & $-$0.05100 $\sigma^x_3 \sigma^x_2 \sigma^y_1 \sigma^y_0$ & $-$0.05711 $\sigma^x_3 \sigma^x_2 \sigma^y_1 \sigma^y_0$  &0.03964 $\sigma^z_2$\\
 \bottomrule
\end{tabular}
\end{table}

In this section, we discussed the small case study of molecular hydrogen, observing that the optimal ordering is described by alternating between fully commuting sets of terms.  We proceed to present two generalized ordering schemes derived from this heuristic, and analyze their performance across a  variety of molecular systems.

\section{Generalized Ordering Strategies}
\label{sec:generalised}
While analysis of the hydrogen molecule yielded interesting results regarding the spread of ordering strategies, it would be desirable to find an ordering strategy that is effective in the general case.  For this, we require an analysis of a variety of systems, and ideally those of a chemically interesting size.  While the hydrogen study does not provide this, the optimal ordering at equilibrium bond length does provide us with a potential starting point for our investigation.

\subsection{Methods}

   To contrast our ordering schemes against other possible alternatives, it is necessary to briefly review the conventions we use to describe other ordering schemes, previously discussed in Reference~\cite{tranterComparisonBravyiKitaev2018}.

Perhaps the most immediately obvious ordering scheme is the \emph{magnitude} ordering.  Here, terms are ordered according to the magnitude of their coefficient, from largest to smallest---the term with the largest magnitude coefficient is simulated first, followed by the second largest magnitude coefficient, and so on.  This ordering scheme has the immediate appeal that high magnitude terms are simulated first.  As a result, one could expect a reduction in the number of high magnitude terms in the error operator.  Loosely, this is due to implementing high magnitude terms first, as opposed to later in the sequence where they may compound earlier errors.  However, this approach does not take into consideration the structure of the error operator---high magnitude terms may not result in a meaningful increase in error if, for example, they commute with many other terms.

The \emph{lexicographic} ordering is an ordering scheme which attempts to maximize the similarity of the Pauli strings of adjacent terms. This is essentially a numerical ordering with respect to the Pauli strings.  While this scheme may seem arbitrary, it is known to result in a maximum amount of gate cancellation, with commensurate reduction in overall quantum computational cost~\cite{hastingsImprovingQuantumAlgorithms2015}.  However, there is little reason to suspect that this ordering would be beneficial for the purposes of Trotter error.  Such an increase could lead to the requirement for a greater number of Trotter steps, undermining its advantages.  It should be noted that our strategy for ordering terms lexicographically is a small modification from that used in other work~\cite{hastingsImprovingQuantumAlgorithms2015}.  Whereas other approaches have grouped the qubit Hamiltonian terms by their fermionic operator, our approach does not maintain this structure, and instead stores the qubit Hamiltonian as a list of individual weighted Pauli string terms.  As such, we do not treat any terms based on their fermionic role.  Similarly, our ordering is based purely on the structure of the individual Pauli strings, rather than the fermionic terms they correspond to.

Testing on random Hamiltonians~\cite{tranterQuantumChemistryQuantum2018} proved relatively optimistic for the prospects of some of the ordering schemes described in later sections of this paper.  However, it has been established that testing on random Hamiltonians does not realistically capture the behavior of real molecular Hamiltonians when considering Trotter errors.  A more rigorous analysis requires the use of real chemical Hamiltonians.  We therefore used a set of 44 molecular Hamiltonians, largely using the same dataset as in Reference~\cite{tranterComparisonBravyiKitaev2018}.  Table~\ref{table:big_systems} shows a breakdown of these systems.  Equilibrium molecular geometries were gathered from the NIST CCCBDB database optimized at the Hartree--Fock level~\cite{johnsoniiiNISTComputationalChemistry2016}.  Molecular orbital integrals in the Hartree--Fock basis were obtained from Psi4~\cite{parrishPsi4OpenSourceElectronic2017} and OpenFermion~\cite{mccleanOpenFermionElectronicStructure2017}.  As in Reference~\cite{tranterComparisonBravyiKitaev2018}, the procedure for determining the exact Trotter error was to directly calculate the expectation value of the Trotterized Hamiltonian with the exact ground state, using our Python code.  The Trotter evolution time for each Hamiltonian was set

\begin{equation}
\label{eqn:trotterTime}
t = 
\begin{cases}
1, & \text{where } \left|E_{FCI}\right| < 2 \pi \\
\frac{1}{2 \pi \left \lfloor{\frac{\left|E_{FCI}\right|}{2 \pi}}\right \rfloor}, & \text{otherwise} 
\end{cases}
\end{equation}

using the full configuration interaction energy for computational convenience.  In a real simulation, the energy provided by a lower level of theory would be used.

\begin{table}[]
\centering
\caption[]{The molecular dataset used.  Note that most of the systems involving a non-minimal basis set were $\ce{H2}$ and $\ce{HeH+}$ systems, as specified in Appendix~\ref{app:systems}.  The polyatomic category includes molecules, ions and radicals.} \label{table:big_systems}
\begin{tabular}{lcccc}
\toprule
\bf{Qubits} &\bf{ 1-10} & \bf{11-20} & \bf{21-30 }& \bf{Total} \\
\midrule
Polyatomic & 1 & 14 & 5 & 20 \\
Atoms & 4 & 6 & 0 & 10 \\
Ions & 2 & 2 & 0 & 4 \\
Other bases & 4 & 4 & 2 & 10 \\
\\
Total & 11 & 26 & 7 & 44 \\
\bottomrule
\end{tabular}
\end{table}

\subsection{Results---Magnitude Ordering}

\begin{figure}[]
	\centering 
	\includegraphics[width=0.8\textwidth]{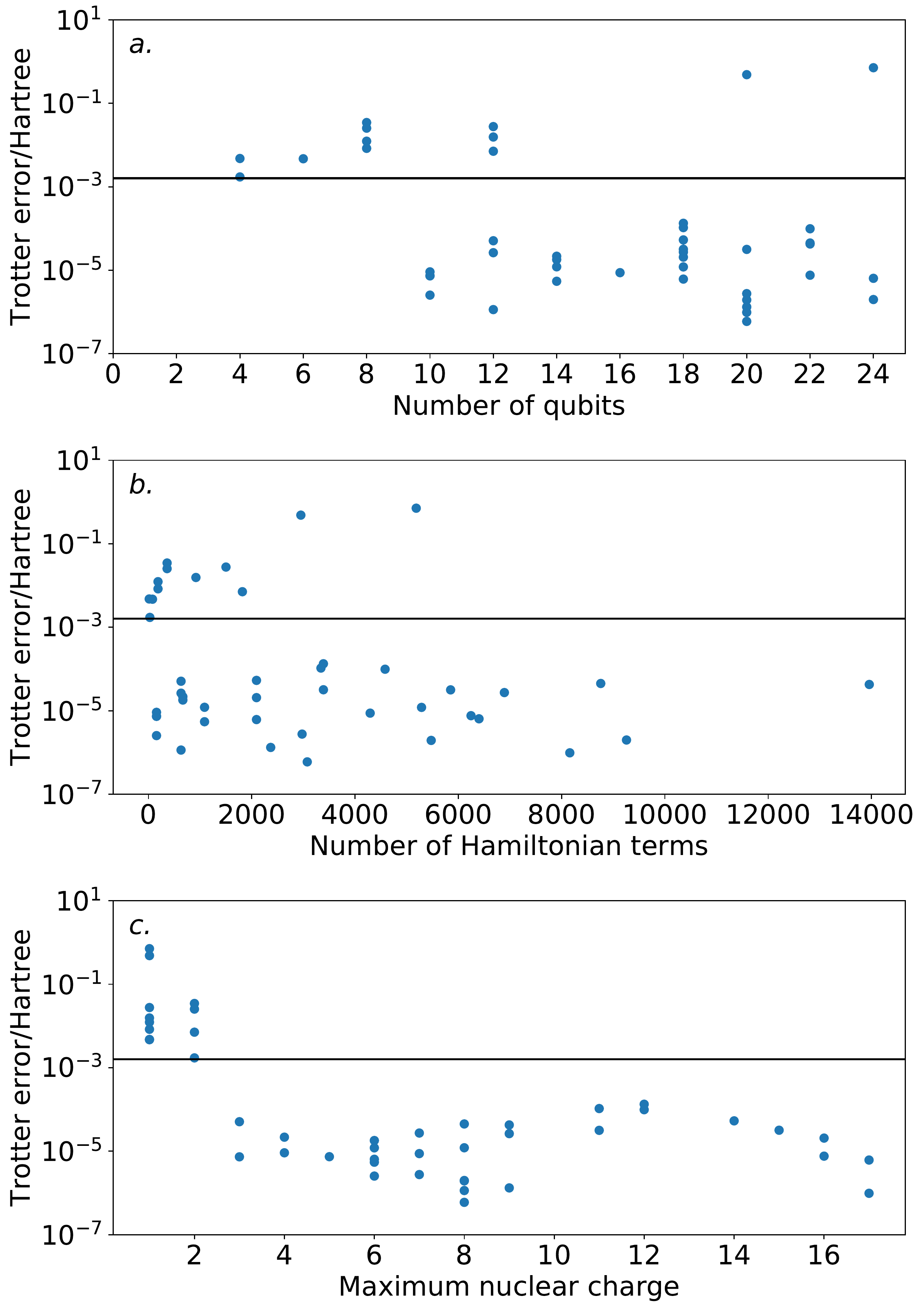}
		\caption[]{Trotter errors for the dataset of molecular Hamiltonians using a magnitude ordering.  The vertical bar indicates chemical accuracy.  Most of the systems achieve chemical accuracy with one Trotter step.  a:  versus the number of spin-orbitals.  Most of the high-error results are for low numbers of spin-orbitals.  b:  versus the number of terms in the Hamiltonian.  Again, most of the high-error results are for low numbers of terms.   c: versus the maximum nuclear charge.  All of the high-error systems are for systems with exclusively light atoms, and the overall trend roughly follows the predictions of prior literature.}
			\label{fig:magnitude}
\end{figure}

To provide some context for our discussion of Trotter ordering strategies, we first consider the Trotter error incurred for our systems using a magnitude ordering.  Figure~\ref{fig:magnitude} demonstrates this against a variety of properties.  The majority of the systems demonstrated Trotter error below the threshold for chemical accuracy.  While there is little indication of increasing Trotter error with the number of spin-orbitals, it should be noted that the systems studied are all within the regime that can be practically simulated on modern classical computers.  It is unclear as to whether this trend will persist beyond this regime.  Nonetheless, this is encouraging for experimental studies in the near future, where the number of available qubits will be highly constrained.

There is no obvious correlation between the Trotter error and either the number of spin-orbitals  or the number of terms in the Hamiltonian.  However, when plotted against the maximum nuclear charge (i.e.,    the nuclear charge of the heaviest atom in the system)---as suggested by the previous work of \citet{babbushChemicalBasisTrotterSuzuki2015}---there is a notable, albeit loose, trend.  All of the systems with single step magnitude ordering Trotter error insufficient for chemical accuracy involve only $\ce{H}$ and $\ce{He}$ nuclei.  This could be affected by the special handling of the Trotter time for low energy systems in Equation    \eqref{eqn:trotterTime}.  The results here reinforce the need for benchmarks of quantum approaches to electronic structure theory problems to consider systems with heavy atoms.

Beyond these systems, there is a peak in Trotter error around a maximum nuclear charge of $11$.  This is in agreement with the results of Reference ~\citet{babbushChemicalBasisTrotterSuzuki2015}, where it is observed that systems with mostly full spin-orbitals will incur low Trotter error.  The peak in our data is due to the presence of $\ce{Na}$ or $\ce{Mg}$, which in an STO-3G basis have many unfilled spin-orbitals.

\subsection{Statistics of Commuting Hamiltonian Subsets}
One approach to using commutativity structure to inform ordering strategy is to consider coloring of the incompatibility graph of the qubit Hamiltonian.  As in Figure~\ref{fig:hyCommutatitivity2}, we can consider this structure by representing the commutativity of the terms in the Hamiltonian as a graph, with terms in the Hamiltonian corresponding to nodes and an edge representing the case where terms do not commute.  Generating such a graph requires $\mathcal{O}\left( N_{terms}^2 \right)$ time, as calculating the commutator of two arbitrary terms is classically efficient.

Following our example of the Hydrogen molecule, a strategy of dividing the Hamiltonian into sets of mutually commuting subsets---sets of terms where all members commute---can be followed.  This is equivalent to finding a coloring of the incompatibility graph.  Unfortunately, it is well known that finding a coloring with a minimal number of colors is NP-hard~\cite{gareySimplifiedNPcompleteProblems1974}.  However, heuristics---often using a greedy approach---have been developed~\cite{kosowskiClassicalColoringGraphs2004}.  There is no immediately obvious reason to suspect that coloring the Hamiltonian using a heuristic is problematic for the purposes of our analysis.  Indeed, these heuristics have seen frequent recent use for partitioning electronic structure Hamiltonians, for the purpose of measurement reduction in variational quantum algorithms~\cite{verteletskyiMeasurementOptimizationVariational2019,izmaylovUnitaryPartitioningApproach2019,crawfordEfficientQuantumMeasurement2019,zhaoMeasurementReductionVariational2019}.  Nonetheless, it does suggest that caution should be used when considering the generalizability of the results presented.  

\begin{figure}[]

	\centering
    {\includegraphics[width=\textwidth]{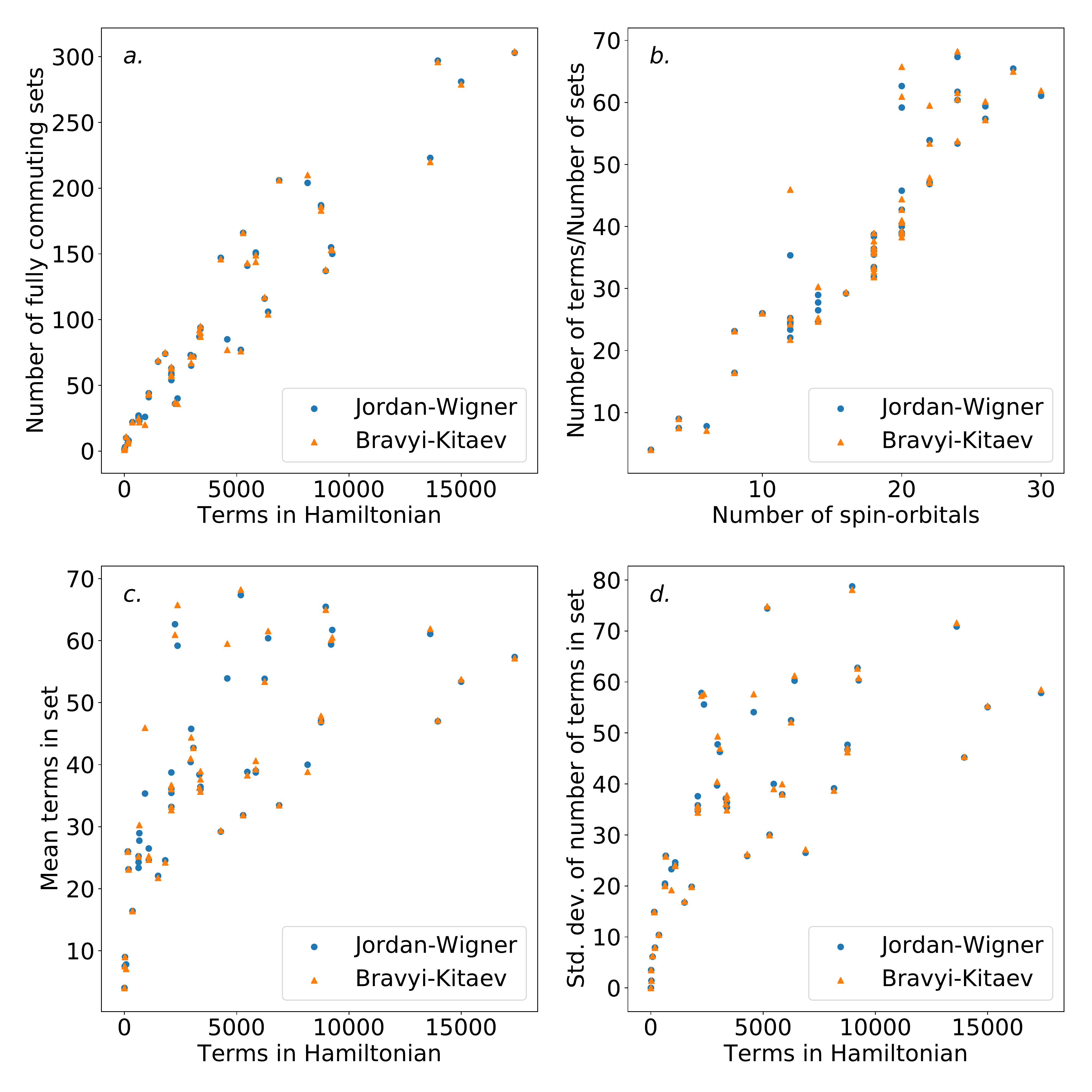}}
        \caption[]{Statistics of the fully commuting sets of terms found in the coloring of the Hamiltonians in the dataset.  a: number of fully commuting sets versus the number of terms in the Hamiltonian. b: number of independent sets divided by the number of terms, versus the number of spin-orbitals.  A roughly linear trend is observed, indicating a $\Theta\left(N^3\right)$ scaling. c: average number of terms in each fully commuting subset for a given Hamiltonian.  d: standard deviation of number of terms in each fully commuting subset for a given Hamiltonian.  The increasing variance in group sizes could be problematic for ordering purposes.}
    \label{fig:numberSets}
\end{figure}

Graphs representing the Hamiltonians in our dataset were generated using the NetworkX Python package~\cite{arica.hagbergProceedingsPythonScience}.  Due to the relative computational ease of this procedure, our dataset here was extended by 16 additional systems, as indicated by Appendix~\ref{app:systems}.  From here, colorings were generated using the greedy coloring method provided by the same package, using an independent set strategy~\cite{kosowskiClassicalColoringGraphs2004}.  Coloring schemes for both Jordan--Wigner and Bravyi--Kitaev Hamiltonians were generated.  Figure~\ref{fig:numberSets} shows the number of independent sets found with regard to both the number of terms in the overall Hamiltonian, and the number of spin-orbitals describing the Hamiltonian.  As is to be expected, the number of sets increases with the number of terms in the Hamiltonian.  

There are $\mathcal{O}(N^4)$ terms in the initial electronic Hamiltonian.  Of these, terms that do not share any indices with each other will clearly commute, even in the absence of the use of the Jordan--Wigner or Bravyi--Kitaev transformations.  As such, it is reasonable to expect the number of independent sets to be $\mathcal{O} (N^3)$.  Therefore the ratio of the number of terms and independent sets found can be expected to scale roughly linearly with the number of spin-orbitals involved.  This scaling is shown in Figure~\ref{fig:numberSets}, outside the smallest Hamiltonians.  Notable outliers are present, likely due to molecular symmetries.  Little difference is observed between the Jordan--Wigner and Bravyi--Kitaev Hamiltonians.  This is to be expected, as the commutativity structure is defined by the physical electronic Hamiltonian; there is no obvious reason to expect that the mapping technique used would significantly affect this.  Our results here are in agreement with those in Reference ~\cite{zhaoMeasurementReductionVariational2019}, where the same scaling was observed for the dual problem (i.e.,    coloring the compatibility graph of the Hamiltonian).

Figure~\ref{fig:numberSets} also demonstrates the average and standard deviation of the size of the independent sets found for each Hamiltonian.  The increased average size of the groups is unsurprising, as the number of terms increases faster than the number of sets.  The increasing variance in the size of the sets may be undesirable.  This is due to the fact that evenly-sized groups could be advantageous for ordering schemes, due to an increased ability to distribute the placement of terms.  It is possible that the use of an alternative coloring strategy could circumvent this, and further work is required to assess whether this has an impact on the ordering schemes presented in Section~\ref{subsec:subsetOrdering}.

\label{subsec:subsetOrdering}

\subsection{Subset-Based Ordering Schemes}

The purpose here of dividing the Hamiltonian into mutually commuting subsets is to examine how a Trotter ordering which takes this into account will perform.  The above analysis of the hydrogen molecule suggested that alternating between commuting sets may be an effective scheme for Trotter ordering.  With Hamiltonians partitioned into commuting subsets as above, this scheme can be extended to larger systems.  In this method, an ordered Hamiltonian is generated by sequentially picking terms from each subset until all subsets are depleted.

Two methodologies for this ordering algorithm were considered.  In the first---the \emph{depleteGroups} strategy---the sets were cycled through, picking the highest magnitude term from each and appending this to the ordered Hamiltonian.  The second---the \emph{equaliseGroups} strategy---at each stage picked the highest magnitude term in the largest subset, appending it to the ordered Hamiltonian.  Where there were multiple ``largest subsets", the highest magnitude term in the union of all largest subsets was used.  Whereas the former strategy ensures that the sets are consistently cycled through until depletion, the latter ensures that the sets are evenly distributed throughout the Trotterized Hamiltonian.

The Trotter error for one Trotter step (using a second-order Trotter approximation) was calculated for each Hamiltonian using both ordering strategies, using the approach discussed in Section~\ref{sec:generalised}.  Clearly, as this includes systems requiring more than $8$ qubits, an exhaustive search of all possible orderings is not possible for all but the smallest of the systems included.  Instead, we compare the depleteGroups and equaliseGroups orderings against other ordering schemes described above.  We consider the performance of each ordering in terms of the number of spin-orbitals considered in the molecular system, and in the maximum nuclear charge.  

\begin{figure}[]
	\centering
    {\includegraphics[width=0.6\textwidth]{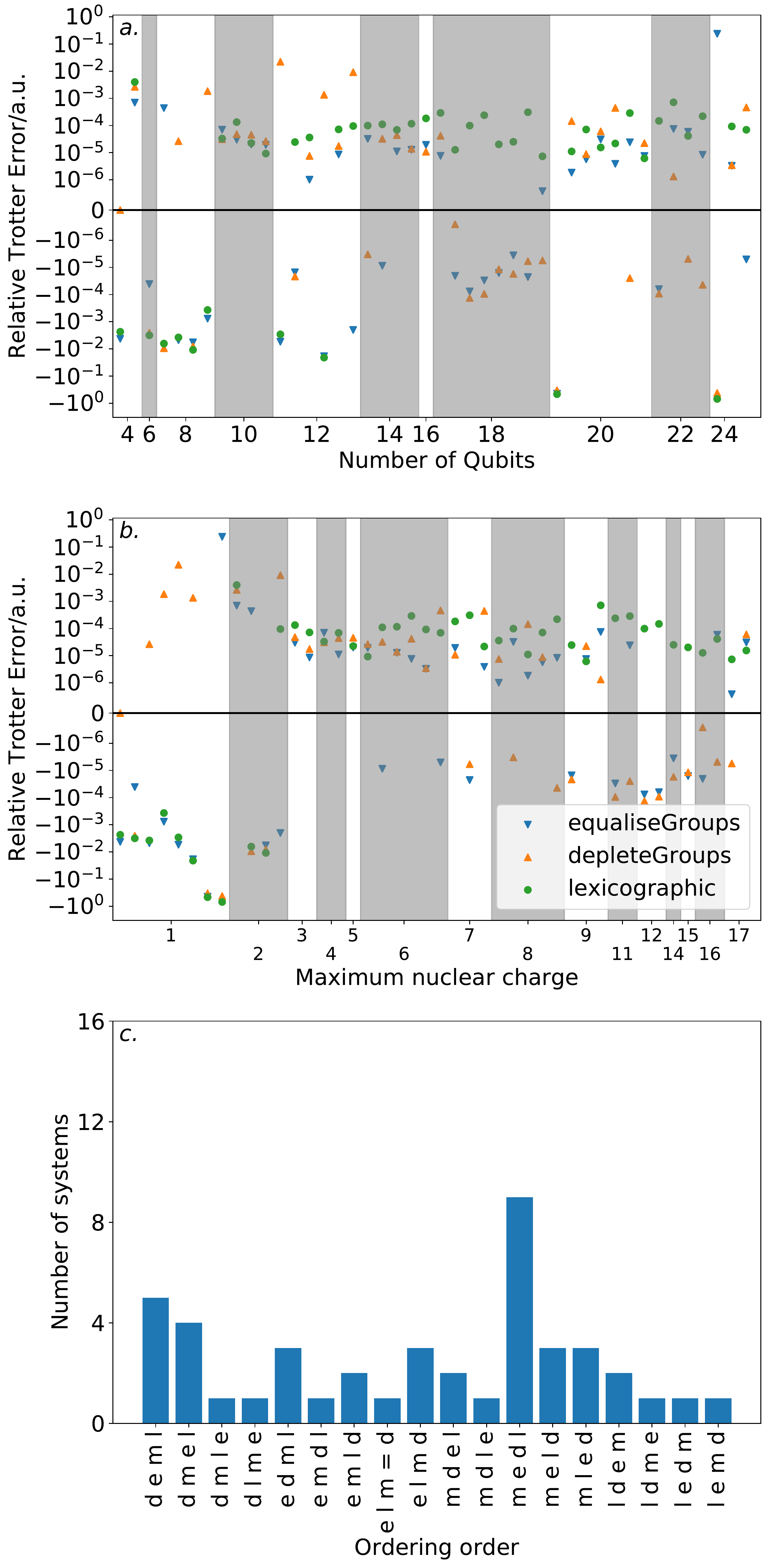}}
    \caption[]{Trotter error of the depleteGroups and equaliseGroups orderings relative to a magnitude ordering.  Upper plots are linear within $\pm 10^{-6}$.  a: by number of qubits.  b: by maximum nuclear charge.  Again, the magnitude ordering is preferable in most cases; however, for systems with period three atoms, the depleteGroups and equaliseGroups are best.   c: frequency of ``ordering orders'', being the sequence of ordering performance.  The distribution here is relatively flat.}
    \label{fig:equaliseDeplete}
\end{figure}

Figure~\ref{fig:equaliseDeplete} shows the results of this analysis.  High variance in ordering performance is associated with systems with only light atoms, but for most systems with heavier atoms, there is only a roughly $0.0001$   a.u. difference in Trotter error between the orderings.  This is a large variance in comparison to the absolute Trotter error.  However, Figure~\ref{fig:magnitude} shows that using a magnitude ordering for these systems yielded an error that is sufficient for chemical accuracy.  The observed difference between orderings is insufficient to increase the error above this threshold.  As such, these results suggest that ordering choice when performing VQE can be determined by other factors, such as circuit length minimization.  This is in sharp contrast to the results in Section~\ref{sec:hydrogen} on the hydrogen molecule in an STO-3G basis, where the Trotter error was dramatically impacted by the Trotter ordering, resulting in a differing amount of Trotter steps required for chemical accuracy.  This reinforces the need to consider larger systems when benchmarking methods.  Conversely, if phase estimation is to be performed, it is likely that this error would be compounded due to applications of higher powers of the Trotterized unitary.  In this context, an optimal ordering scheme in terms of Trotter error would be more necessary.

The equaliseGroups and depleteGroups strategies appear reasonably promising.  The depleteGroups scheme is superior to the equaliseGroups strategy in all bar one of the systems with more than 20 spin-orbitals, and performs roughly commensurately elsewhere.  We conclude that of the depleteGroups and equaliseGroups strategies, the depleteGroups strategy should be favored, although it is possible that this is a result of the uneven group sizes shown by Figure~\ref{fig:numberSets}.  An inspection of Figure~\ref{fig:equaliseDeplete} shows that the depleteGroups strategy is better than the magnitude ordering in eighteen cases.

More consistency is observed with respect to the maximum nuclear charge.  Highly variable performance is observed with light atoms; however beyond this, the difference in Trotter error is relatively minor.   For several of the systems consisting of the most (22) spin-orbitals, the depleteGroups strategy begins to dramatically outperform all other orderings considered---including the magnitude ordering.  For systems involving period three  atoms, there is only one exception to this result.  More data  are required to assess whether this trend consistently extends to other, and larger, systems.

It should be noted that the process reported here to color the Hamiltonian commutativity graph was relatively simple, examining only one possible ordering strategy using a standard library.  Future work could investigate the impact of altering the details of this scheme.  A variety of factors could be considered here.  For example, it is not immediately obvious whether it would be preferable to split the graph into few large independent sets, or many smaller ones.  As mentioned above, various schemes of Hamiltonian term partitioning have been developed to reduce the cost of variational quantum algorithms, by combining sets of commuting or anticommuting terms~\cite{verteletskyiMeasurementOptimizationVariational2019,izmaylovUnitaryPartitioningApproach2019,crawfordEfficientQuantumMeasurement2019,zhaoMeasurementReductionVariational2019}.  It remains an open question as to whether Trotter ordering schemes based on graph coloring heuristics can be applied to Hamiltonians with terms combined in such a fashion.
 
The results of the depleteGroups strategy are encouraging, although they do not achieve improvement over a magnitude ordering in all cases.  Further work examining larger systems is required to test whether the observed improvement with the depleteGroups strategy is maintained at larger numbers of spin-orbitals, although this may prove computationally difficult.

In complement to the above discussion, an entirely different approach to ordering could be considered, which relies upon the Trotter error operator.  As the norm of this can be efficiently (albeit slowly) determined classically, it is possible that this information could be used to directly inform an ordering strategy---rather than indirectly through the commutation relations of terms.  For completeness, we consider such an approach now.

\section{Error Operator Strategies}
\label{sec:errorOperator}
\subsection{Small Systems}

As discussed above, the Trotter error operator can be used to determine the Trotter error of a given Trotterization scheme, with the norm of this operator giving a (loose) upper bound on the Trotter error.  This information could be used to guide an ordering strategy. 

To this end, we first examine how the Trotter error operator norm varies with respect to the Trotter ordering.  The Trotter error operator norm of each ordering for the hydrogen molecule in a minimal basis was calculated, using a second-order Trotter approximation, with the fully commuting terms removed as in Section~\ref{sec:hydrogen}.    Similarly, the Trotter error operator norm for 100,000 randomly chosen orderings of the helium hydride ion---also in a minimal basis---was calculated, along with the expectation of the Trotterized unitary with the exact ground state for each.  As there are $27! \approx 10^{28}$ possible orderings for this Hamiltonian, our sample is only a small fraction of the entire ordering~space.

\begin{figure}[]
	\centering
    {\includegraphics[width=0.7\columnwidth]{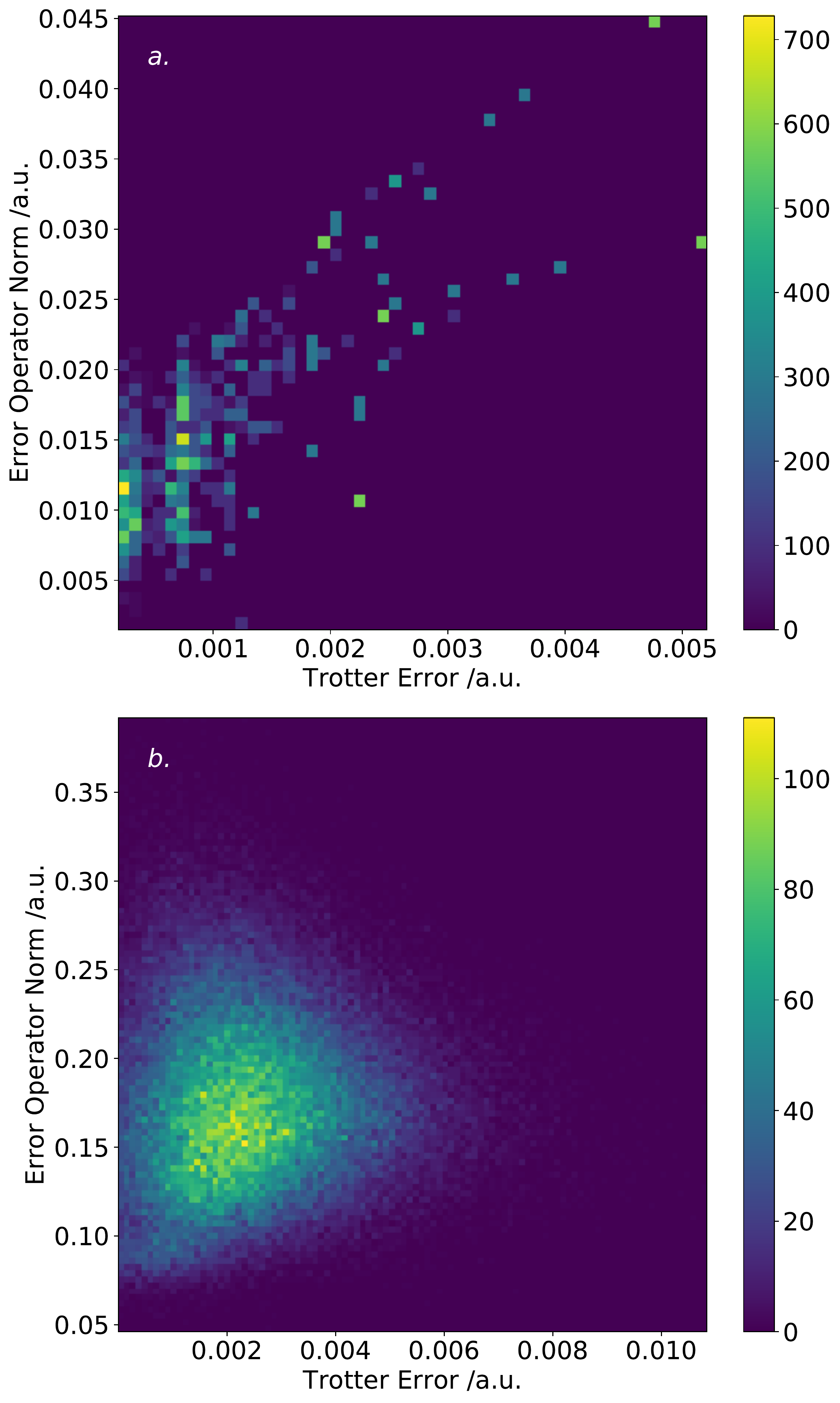}}
        \caption[]{The Trotter error operator norm versus true Trotter error, for various orderings.  a: hydrogen molecule in a minimal basis.  Two hundred and fifty bins are used.  Loose correlation is observed, although there is ambiguity for a true Trotter error of less than 0.001 a.u.  b: helium hydride in a minimal basis, using 100,000 random orderings.  One thousand  bins are used.  Little obvious correlation is observed.}
    \label{fig:errorOp1}
\end{figure}

The results of this analysis are shown in Figure~\ref{fig:errorOp1}.  The two systems display different trends.  For the hydrogen molecule, there is a loose correlation between the Trotter error operator norm and the the true Trotter error.  The worst possible orderings---incurring an approximate factor of 5 increase in Trotter error versus the best orderings---also obtain a low error operator norm.  This is encouraging, as it suggests that relying on the Trotter error operator norm will not result in an extremely poor ordering, at least for the hydrogen molecule in a STO-3G basis.  However, it should be noted that within the low Trotter error region (below 0.001 a.u.), there is a broad spread of error operator norms, indicating that this approach may not be effective for finding truly optimal orderings.  The helium hydride results are less clear.  Indeed, the worst orderings tested in terms of true Trotter error have a relatively low error operator norm.  Nonetheless, in both cases, the lowest true Trotter error orderings also resulted in a low Trotter error operator norm.  As for the hydrogen molecule shown here and discussed above, the helium hydride results show a high density of orderings in the low Trotter error region---suggesting that most ordering schemes will be ``good enough ``, provided they do not systematically result in falling into the high Trotter error region.

\subsection{Term Insertion Error Operator}

While the Trotter error operator can be determined in polynomial time on a classical computer, the triple sum of triple commutators present in Equation    \eqref{errorOperator} nonetheless presents some computational difficulty.  An examination of the Trotter error operator suggests that it can be calculated in $\mathcal{O}(N_{terms}^3)$ time.  Assuming that there are $\mathcal{O}(N_o^4)$ terms in the Hamiltonian (where $N_o$ is the number of spin-orbitals involved), the calculation of each Trotter error operator requires $\mathcal{O}(N_o^{12})$ time.  Where the Trotter error operator is calculated a constant number of times, this is surmountable for low numbers of spin-orbitals, due to the sparsity of the coefficients in the electronic Hamiltonian.  However, an ordering scheme based on the calculation of the Trotter error operator may require repeated calculation of varying Hamiltonians.  If the error operator must be evaluated an average of $\frac{N_{terms}}{2}$ times for each of $N_{terms}$ terms, the strategy would require time that scales roughly as the twentieth power of the number of orbitals.  This is somewhat problematic.

In comparing the Trotter error operator of two orderings, it is therefore preferable to consider the difference between their error operators, thereby reducing the triple sum to a double sum.  For clarity, we define the term operator

\begin{equation}
C_{\alpha \beta \gamma} = \frac{{-\Delta_t}^2}{12} \left[H_\alpha \left(1 - \frac{\delta_{\alpha,\beta}}{2}\right),\left[H_\beta,H_\gamma \right] \right] \,.
\end{equation}

If we insert a term $H_i$ at index $i$ into a given Trotterized Hamiltonian, and divide the $\beta$ sum into $\beta < i$, $\beta = i$ and $\beta > i$ components, the new Trotter error operator $V_i$ is then:

\begin{equation}
V_i = \sum_{\alpha \leq \beta} \sum_{\beta < i} \sum_{\gamma < \beta} C_{\alpha \beta \gamma} + \sum_{\alpha \leq i} \sum_{\gamma < i} C_{\alpha i \gamma} + \sum_{\alpha \leq \beta} \sum_{\beta > i} \sum_{\gamma < \beta} C_{\alpha \beta \gamma} .
\end{equation}

We separate the terms where $\alpha = i$ or $\gamma = i$ from the final sum to yield:

\begin{equation}
\label{eqn:vi}
V_i = \sum_{\alpha \leq \beta} \sum_{\beta < i} \sum_{\gamma < \beta} C_{\alpha \beta \gamma} + \sum_{\alpha \leq i} \sum_{\gamma < i} C_{\alpha i \gamma} + \sum_{\beta > i} \left(\sum_{\substack{\alpha \leq \beta \\ \alpha \neq i}} \sum_{\substack{\gamma < \beta \\ \gamma \neq i}}
C_{\alpha \beta \gamma} + \sum_{\substack{\alpha \leq \beta \\ \alpha \neq i}} C_{\alpha \beta i} + \sum_{\gamma < \beta} C_{i \beta \gamma} \right)
\end{equation}

The difference between $V_i$ and $V$---the terms added to the error operator by the insertion of term $H_i$ at index $i$---will be the terms that contain either $C_{i \beta \gamma}$, $C_{\alpha i \gamma}$ or $C_{\alpha \beta i}$.  This is true of all terms in the single and double sums in Equation   \eqref{eqn:vi}, and none of the terms in the triple sums.  Replacing $C_{\alpha \beta \gamma}$ we therefore have 

\begin{align}
\label{eqn:termInsertion}
V'_{i} &= \frac{{-\Delta_t}^2}{12} \left( \sum_{\alpha \leq i} \sum_{\gamma < i} \left[H_\alpha \left(1 - \frac{\delta_{\alpha,i}}{2}\right),\left[H_i,H_\gamma \right] \right] \right. \nonumber \\ &\left. + \sum_{\beta > i} \left( \sum_{\gamma < \beta} \left[H_i ,\left[H_\beta,H_\gamma \right] \right] + \sum_{\substack{\alpha \leq \beta \\ \alpha \neq i}} \left[H_\alpha \left(1 - \frac{\delta_{\alpha,\beta}}{2}\right),\left[H_\beta,H_i \right] \right]\right)\right)
\end{align}

the Trotter term insertion error operator, which consists of the terms added to the error operator upon inserting a term $H_i$ at index $i$.  Provided a reference error operator, this is substantially easier to compute than the calculation of the entire error operator afresh.

\subsection{Error Operator Based Ordering Schemes}
\begin{figure}[]
	\centering
    {\includegraphics[width=0.55\textwidth]{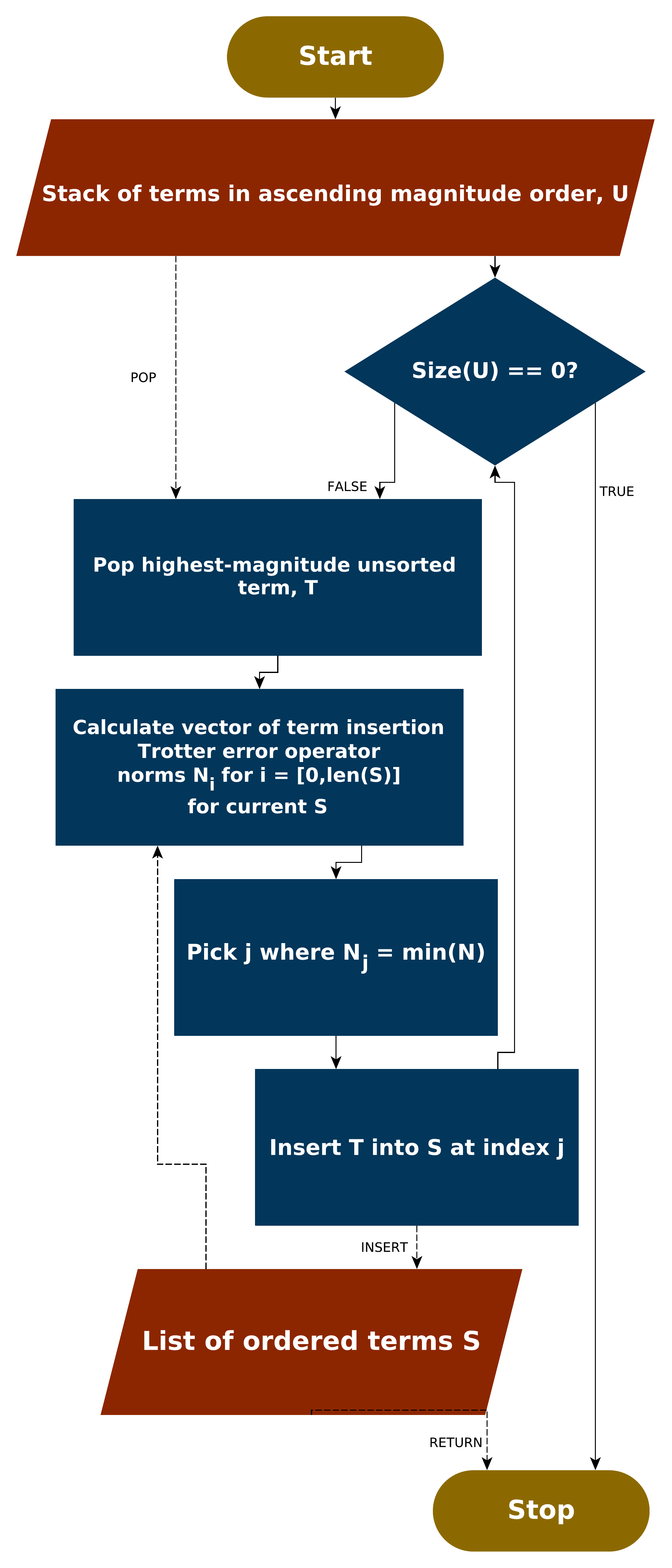}}
        \caption[]{The procedure for performing the error operator norm minimization ordering.}
        \label{fig:errorOpFlowchart}
\end{figure}

We can now define a Trotter error operator based ordering scheme, following the logic of an insertion sort.  We begin with the unsorted Hamiltonian.  The highest magnitude term is then removed from the unsorted Hamiltonian, and forms the first term in the sorted Hamiltonian.  The Trotter term insertion error operator (Equation    \eqref{eqn:termInsertion}) is calculated in the cases where the second term is inserted either before or after the initial term.  The norm of this operator is calculated in both cases.  The scenario wherein the term insertion Trotter error operator norm is minimized is chosen as the correct branch, and the term is placed in the corresponding location.  The third highest magnitude term is then tested in each of the three new potential locations and placed in the one which minimizes the term insertion Trotter error operator norm.  This process is repeated until the unsorted Hamiltonian is depleted.  The procedure for ordering the Hamiltonian in this manner is depicted as Figure~\ref{fig:errorOpFlowchart}.

Despite eliminating the need to calculate the Trotter error operator at every step in the optimization, the procedure remains extremely slow, requiring $\mathcal{O}\left(N_{terms}^4\right)$ time.  As such, this ordering scheme was applied to a subset of $36$ of the Hamiltonians in our dataset, and the one-step Trotter error calculated.  Appendix~\ref{app:systems} shows which systems are included in this subset.

\begin{figure}[]
	\centering
    {\includegraphics[width=0.68\textwidth]{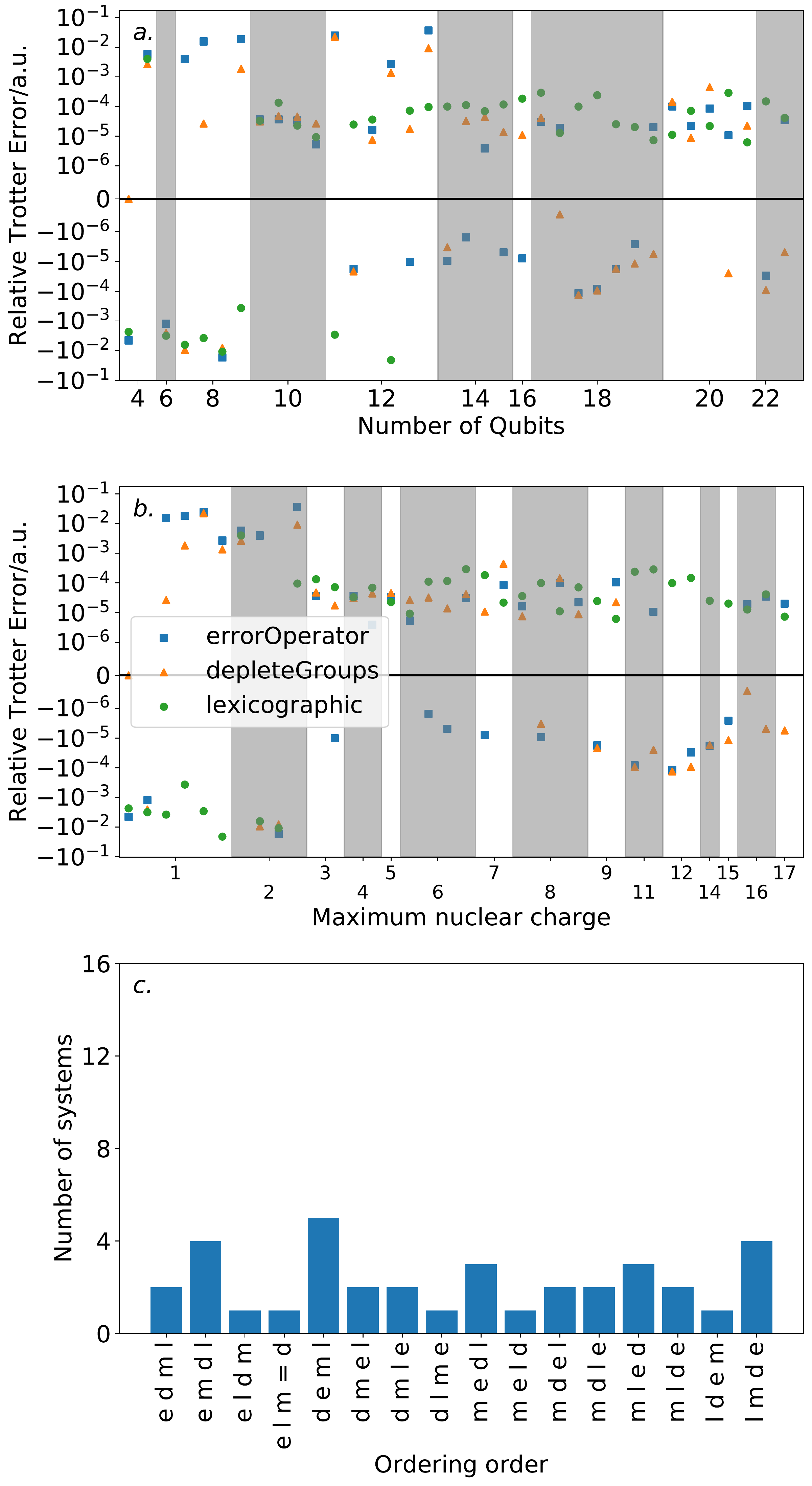}}
        \caption[]{Trotter error of the errorOperator ordering relative to a magnitude ordering.  Upper plots are linear between $\pm 10^{-6}$.  a: by number of qubits.  b: by maximum nuclear charge.  As with the previous orderings, the variance between Trotter ordering schemes is low for systems involving heavy atoms.  In these cases, the errorOperator ordering performs roughly commensurately with the magnitude ordering. c: frequency of ``ordering orders'', being the sequence of ordering performance.  The distribution here is relatively flat.}
    \label{fig:errorOperatorNumerics}
\end{figure}

Figure~\ref{fig:errorOperatorNumerics} shows the results of this analysis.  Again, for large numbers of qubits or the inclusion of heavy atoms, the difference in Trotter error between the orderings is relatively minor.  For systems with more than 12 qubits, the errorOperator ordering results in an approximately $\pm 10^{-4}$--$\pm 10^{-5}$ a.u.  difference in error relative to a magnitude ordering.  This is similar to that observed for the other ordering schemes considered.  While outperforming the magnitude ordering in around half of the systems with maximum nuclear charge greater than $6$, it does not consistently reach the performance of the depleteGroups ordering.  Two possible inferences could be drawn from this.  Firstly, the greedy optimization heuristic used to minimize the Trotter error operator norm may be at fault.  This is particularly evidenced by the decreased performance of the errorOperator ordering for systems with 20 or more spin-orbitals, due to the large problem space.  Alternatively, it could be simply due to the Trotter error operator norm being a loose bound on the actual Trotter error.  While it is regrettable that this ordering does not appear to systematically outperform the other orderings considered, these factors---coupled with the computational difficulty of repeated calculations of the Trotter error operator---suggest that it will be difficult to find an effective and scalable ordering strategy reliant upon the Trotter error~operator.

In several systems, most notably within those involving 18 spin-orbitals, the depleteGroups and errorOperator strategies result in extremely similar errors---a similarity which is not observed for the other ordering strategies.  This indicates some replicated behavior, although further work is required to determine what this may be.

Other work has defined a term importance metric by considering the impact of a given fermionic term on the Hartree--Fock ground state~\cite{poulinTrotterStepSize2015,hastingsImprovingQuantumAlgorithms2015}.     To prevent dependence on the accuracy of the Hartree--Fock state, we do not consider this approach here.  However, given the performance of the error operator ordering compared to a magnitude ordering, it is likely that such an approach could yield improvement with regard to the Trotter error incurred.  Future work could aim to assess whether this is indeed the case.

In this section, we   present  an ordering scheme based on minimizing the Trotter error operator norm.  The approach did not yield consistent improvement upon the depleteGroups strategy, which---in addition to the high computational difficulty of the task---suggests that approaches to ordering by use of the Trotter error operator norm may prove difficult.

\section{Discussion and Conclusions}

In this paper, we discussed ordering schemes for Trotterization from a variety of perspectives.  Brute-force analysis on the STO-3G molecular hydrogen Hamiltonian  and inspection of the Trotter error operator  show  that the commutativity structure of the Hamiltonian is of vital importance in determining the Trotter error incurred by a given ordering.  The commutativity structure---represented by the incompatibility graph---can be efficiently found on a classical computer.  Greedy coloring heuristics can be used to partition this graph into commuting or anticommuting sets, a technique which has been used recently in the context of term reduction in variational quantum algorithms~\cite{verteletskyiMeasurementOptimizationVariational2019,izmaylovUnitaryPartitioningApproach2019,crawfordEfficientQuantumMeasurement2019,zhaoMeasurementReductionVariational2019}. 

  It remains an open question as to how these term reduction techniques can be related to Trotter ordering strategies when implementing variational quantum algorithms.

We reported two ordering schemes dependent on the term commutativity structure, which operate by partitioning the Hamiltonian into totally commuting sets of terms.  Applying a greedy coloring heuristic results in a number of sets that scales approximately as $\mathcal{O}(n^3)$. By using these groups to inform placement of terms, this approach led to the best ordering scheme studied---the depleteGroups strategy.  This strategy had the lowest Trotter error of all orderings in a plurality of Hamiltonians, although for systems involving heavy atoms, the difference when compared to a magnitude ordering was below the threshold for chemical accuracy.

Finally, we considered the use of the Trotter error operator norm to guide ordering strategies.  A greedy algorithm aiming to minimize the norm of the Trotter error operator achieved performance consistent with magnitude ordering, though did not improve upon it in most cases.  However, analysis of the Trotter error operator norm for very small molecules suggested that there is a high density of orderings which incur relatively low---if suboptimal---Trotter error.

Although the depleteGroups strategy showed promise, our analysis has indicated that finding an optimal Trotter ordering is a difficult task, due to the vast space of possible orderings.  However, two results are of interest. 
Relative to the Trotter error, the difference between orderings can be substantial---in several cases, the difference in Trotter error incurred by the best and worst orderings was over two orders of magnitude.  However, for the majority of systems studied, the overall Trotter error is extremely low.  For all except four of the systems with more than 12 spin-orbitals, an ordering could be found that incurred a Trotter error below the threshold for chemical accuracy.  Particularly for variational quantum algorithms---where Trotter decompositions are used to implement a variational Ansatz---this implies that Trotter error will not dominate the computational difficulty of simulations in the NISQ era. 

Determining exact Trotter error on a classical computer is exponentially hard, due to the need to find an exact ground state.  Therefore, there are two ways to scale examinations of Trotter error to systems involving larger numbers of spin-orbitals.  The first is the use of efficient classical heuristics to compute \emph {approximate} estimates of the Trotter error which outperform the Trotter error operator norm.  In particular, the development of classical approximations which reproduce the relative performance of differing Trotter ordering strategies would be a useful direction for future work.  The second---similar to other techniques using a quantum device to improve the quantum algorithm~\cite{luEnhancingQuantumControl2017}---would be the use of NISQ devices themselves to directly evaluate ordering strategies through determination of exact Trotter errors in bulk.  It is likely that a combination of each of these approaches will be necessary in order to determine ordering schemes that minimize Trotter error for large systems.

\begin{acknowledgments}
A.T. is grateful to Michael Bearpark and Seth Lloyd for useful discussions.  This research was funded by EPSRC through Imperial College London's Centre for Doctoral Training in Controlled Quantum Dynamics, grant number EP/G037043/1.  It was also supported by the NSF STAQ project PHY-1818914.  It was further supported by EPSRC grant number EP/L00030X/1.  Additionally, this material is based upon work supported by the Under Secretary of Defense for Research and Engineering under Air Force Contract No. FA8702-15-D-0001. Any opinions, findings, conclusions or recommendations expressed in this material are those of the author(s) and do not necessarily reflect the views of the Under Secretary of Defense for Research and Engineering.
\end{acknowledgments}

\section*{Appendices}
\appendix
\section{Alternative Ordering Schemes}
\subsection{Commutator Ordering}
In this appendix, we discuss an additional two ordering schemes based on the commutativity structure of the Hamiltonian terms.  We observe  in Section~\ref{sec:hydrogen} that the optimal ordering strategy for molecular hydrogen at equilibrium bond length was to divide the Hamiltonian into two mutually commuting subsets, and intersperse them so as to minimize the number of non-zero terms in the error operator, ordering each subset by magnitude.  We present a general ordering scheme based on this approach in Section~\ref{sec:generalised}.  However, generalizing this approach to larger systems is more difficult, as many decompositions of the Hamiltonian into mutually commuting subsets will be possible.  For this reason, we now consider simpler ordering schemes.

This scheme---the \emph{commutator} ordering---is still inspired by the optimal hydrogen ordering at equilibrium bond length.  However, rather than attempting to find specific sets, it instead attempts to order by minimizing the amount of commutativity in the ordered Hamiltonian at each step.

We begin by ordering the Hamiltonian by coefficient magnitude.  The highest magnitude term is then assigned to be the first term in the ordered Hamiltonian.  We then consider the terms that do not commute with this, and the highest magnitude of these is appended to the ordered Hamiltonian.  We keep track of the number of terms in the ordered Hamiltonian with which each term in the unordered Hamiltonian commutes, and we consider the subset of unordered Hamiltonian terms that minimize this number.  Of these, the term with highest magnitude is chosen as the next term in the ordered Hamiltonian.  This process is repeated until the unordered Hamiltonian is exhausted.    Figure~\ref{fig:orderings}~(a) demonstrates this process diagrammatically.

It should be noted that this process does not reproduce the exact optimal strategy for the molecular hydrogen test case, as in this instance the lower magnitude XY-set instead begins the ordering process.  Nonetheless, it does produce an ordering which performs almost as well.

\subsection{Reverse Commutator Ordering}
\begin{figure}[]

	\centering
	\subfloat[]{\includegraphics[width=0.45\textwidth]{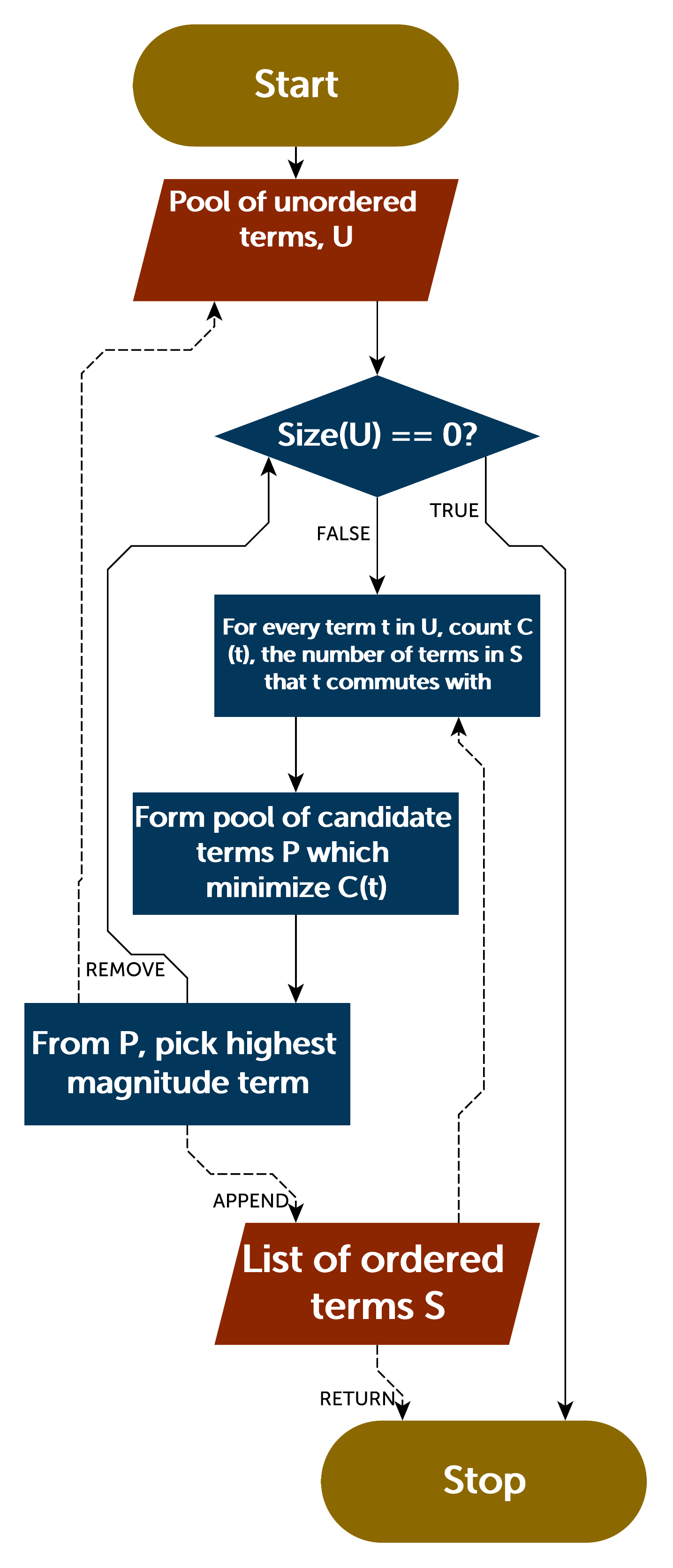}\label{fig:orderingFlowchart1}}
	\subfloat[]{\includegraphics[width=0.45\textwidth]{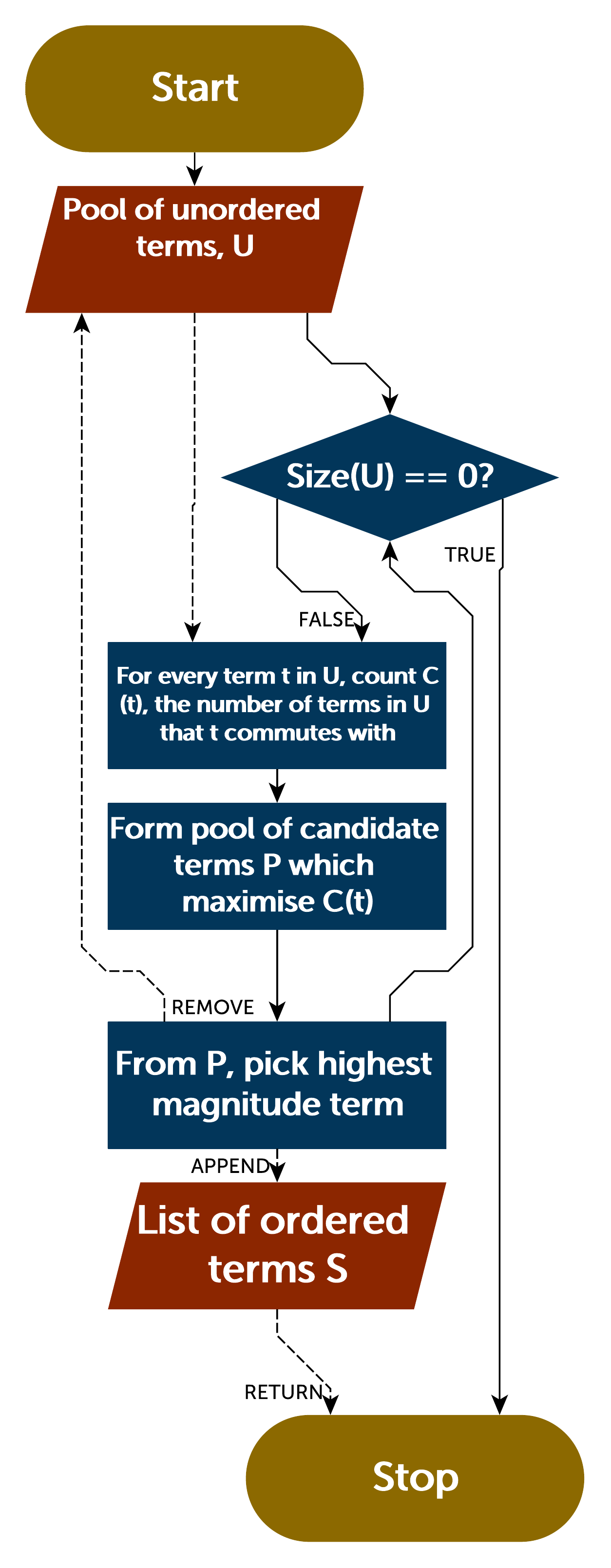}\label{fig:orderingFlowchart2}}
    \caption[Flowcharts representing the commutator and reverseCommutator ordering schemes]{Flowcharts representing the commutator (\textbf{a}) and reverseCommutator (\textbf{b}) ordering schemes.}
    		\label{fig:orderings}
\end{figure}

In addition to this scheme, a slightly modified scheme was considered.  In this scheme---the \emph{reverseCommutator} ordering---we again start with an unordered Hamiltonian comprising of a list of terms.  At each stage, instead of counting the number of terms in the ordered Hamiltonian that the candidate terms commute with, we count the number of terms in the \emph{unordered} Hamiltonian that the candidate terms commute with.  We then pick the terms that \emph{maximize} this number, and append the highest magnitude of these.  This process is depicted in   Figure~\ref{fig:orderings}~(b).

Both  algorithms are essentially greedy.  The commutator ordering at each stage attempts to maximize the non-commutativity of the ordered Hamiltonian, whereas the reverseCommutator ordering at each stage attempts to minimize the non-commutativity of the unordered Hamiltonian.

\subsection{Performance of Commutator and ReverseCommutator Orderings}

\begin{figure}[]
	\centering
    {\includegraphics[width=\textwidth]{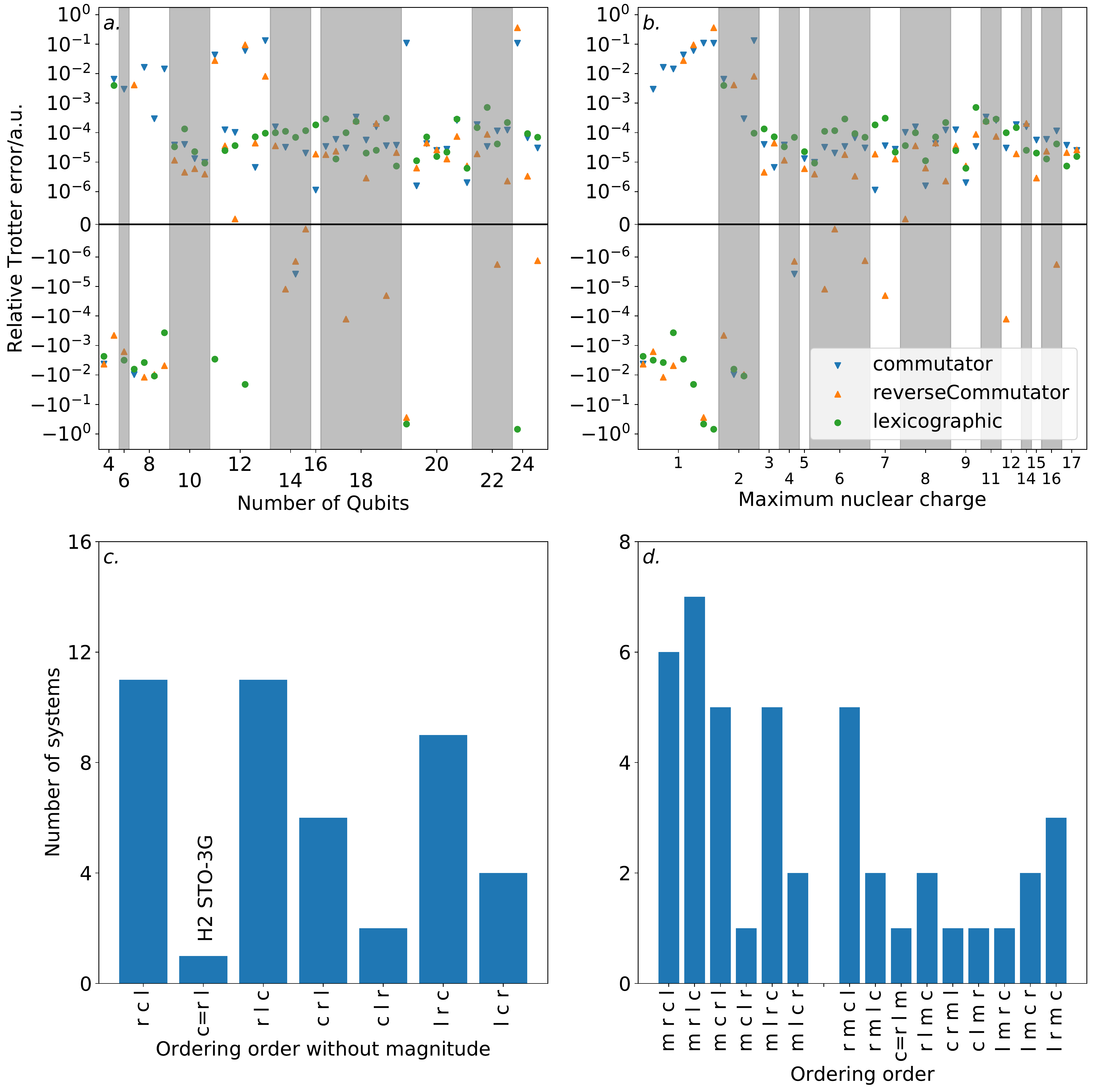}}
    \caption[]{Trotter error of the commutator and reverseCommutator ordering relative to a magnitude ordering.  Upper plots are linear within $\pm 10^{-6}$. a: by number of qubits.  b: by maximum nuclear charge.  The reverseCommutator ordering is best for almost all systems including heavy atoms, however in almost all cases, a simple magnitude ordering outperforms all orderings considered.    c: Frequency of ``ordering orders'', being the sequence of ordering performance, ignoring the magnitude ordering.  d: as lower left, but with the magnitude ordering.} 
    \label{fig:newData1}
    
\end{figure}

Figure~\ref{fig:newData1} demonstrates the results of these simulations.  We first note that, in most cases, the ordering schemes did not outperform a magnitude ordering.  Versus the number of spin-orbitals, the commutator and reverseCommutator orderings do not fare well.  While the reverseCommutator ordering outperformed other orderings by several orders of magnitude in some cases, the improvement is not consistent.  However, assessing the orderings against the maximum nuclear charge indicates that in all bar two instances involving heavy atoms, the reverseCommutator ordering is the best of the three non-magnitude orderings considered.  This suggests that, while the magnitude ordering (or the depleteGroups ordering discussed in the main text) should be preferred for the reduction of Trotter error, there remains potential for ordering schemes based on commutativity structure.

The reliability of a magnitude ordering scheme does suggest improvement here.  Our schemes used the magnitude of terms as only a tie-breaker in cases where candidate terms yielded a common level of non-commutativity against the ordered or unordered terms.  The fact that the magnitude ordering yielded comparatively low error in most cases emphasizes the importance of the magnitude of the resultant terms in the error operator, rather than their mere presence.  Our results here confirm the need for a more sophisticated effort at ordering selection, if minimization of Trotter error is desired.  Additionally, the vast difference in ordering behavior between systems with light and heavy atoms reinforces the need to look at systems beyond STO-3G molecular hydrogen.
\clearpage
\section{Systems in Dataset}
\label{app:systems}
\begin{longtable}{lrrlr}
\caption{The systems included in the dataset.  Geometries were obtained from the NIST CCBDB database~\cite{johnsoniiiNISTComputationalChemistry2016}, and molecular orbital integrals in the Hartee--Fock basis obtained from Psi4~\cite{parrishPsi4OpenSourceElectronic2017} and OpenFermion~\cite{mccleanOpenFermionElectronicStructure2017}.} \\
\toprule
\bf{System} &\bf{ Multiplicity} & \bf{Charge} &\bf{ Basis} &\bf{ Qubits} \\
\midrule 
\ce{B1} & 2 & 0 & STO-3G & 10 \\
\ce{Be1} & 1 & 0 & STO-3G & 10 \\
\ce{C1O1} & 1 & 0 & STO-3G & 20 \\
\ce{C1} & 3 & 0 & STO-3G & 10 \\
\ce{Cl1} & 2 & 0 & STO-3G & 18 \\
\ce{F2} & 1 & 0 & STO-3G & 20 \\
\ce{H1Cl1} \footnotemark[1] & 1 & 0 & STO-3G & 20 \\
\ce{H1F1}\footnotemark[1] & 1 & 0 & 3-21G & 22 \\
\ce{H1F1} & 1 & 0 & STO-3G & 12 \\
\ce{H1He1} & 1 & +1 & 3-21G & 8 \\
\ce{H1He1} & 1 & +1 & 6-311G & 12 \\*
\ce{H1He1} & 1 & +1 & 6-31G & 8 \\*
\ce{H1He1} & 1 & +1 & STO-3G & 4 \\*
\ce{H1Li1O1}\footnotemark[1] & 1 & 0 & STO-3G & 22 \\*
\ce{H1Li1} & 1 & 0 & STO-3G & 12 \\*
\ce{H1Na1} & 1 & 0 & STO-3G & 20 \\*
\ce{H1O1} & 1 & -1 & STO-3G & 12 \\*
\ce{H2Be1} & 1 & 0 & STO-3G & 14 \\*
\ce{H2C1O1}\footnotemark[1] & 1 & 0 & STO-3G & 24 \\
\ce{H2C1} & 3 & 0 & STO-3G & 14 \\
\ce{H2C1} & 3 & 0 & STO-3G & 14 \\
\ce{H2C2}\footnotemark[1] & 1 & 0 & STO-3G & 24 \\
\ce{H2Mg1} & 1 & 0 & STO-3G & 22 \\*
\ce{H2O1} & 1 & 0 & STO-3G & 14 \\*
\ce{H2S1} & 1 & 0 & STO-3G & 22 \\*
\ce{H2} & 1 & 0 & 3-21G & 8 \\*
\ce{H2}\footnotemark[1] & 1 & 0 & 6-311G** & 24 \\*
\ce{H2} & 1 & 0 & 6-311G & 12 \\*
\ce{H2} & 1 & 0 & 6-31G** & 20 \\*
\ce{H2} & 1 & 0 & 6-31G & 8 \\*
\ce{H2} & 1 & 0 & STO-3G & 4 \\*
\ce{H3N1} & 1 & 0 & STO-3G & 16 \\*
\ce{H3} & 1 & +1 & 3-21G & 12 \\
\ce{H3} & 1 & +1 & STO-3G & 6 \\
\ce{H4C1} & 1 & 0 & STO-3G & 18 \\
\ce{H4N1}\footnotemark[1] & 1 & +1 & STO-3G & 18 \\
\ce{Li1} & 2 & 0 & STO-3G & 10 \\
\ce{Mg1} & 1 & 0 & STO-3G & 18 \\
\ce{N2} & 1 & 0 & STO-3G & 20 \\
\ce{Na1} & 2 & 0 & STO-3G & 18 \\
\ce{O2} & 3 & 0 & STO-3G & 20 \\
\ce{P1} & 4 & 0 & STO-3G & 18 \\
\ce{S1} & 3 & 0 & STO-3G & 18 \\
\ce{Si1} & 3 & 0 & STO-3G & 18 \\
\bottomrule

\end{longtable}

\footnotetext[1]{No error operator ordering}
\begin{table}[]
\centering
\caption{The additional systems used for examining the statistics of fully commuting sets of terms in Section~\ref{sec:generalised}.  Geometries were obtained from the NIST CCBDB database~\cite{johnsoniiiNISTComputationalChemistry2016}, and molecular orbital integrals in the Hartee--Fock basis obtained from Psi4~\cite{parrishPsi4OpenSourceElectronic2017} and OpenFermion~\cite{mccleanOpenFermionElectronicStructure2017}.} 
\begin{tabular}[t]{lrrlr}
\toprule
\bf{System} &\bf {Multiplicity} & \bf{Charge} & \bf{Basis} & \bf{Qubits} \\
\midrule 
\ce{Ar1} & 1 & 0 & STO-3G & 18 \\
\ce{C1O2} & 1 & 0 & STO-3G & 30 \\
\ce{Cl1} & 1 & -1 & STO-3G & 18 \\
\ce{F1} & 2 & 0 & STO-3G & 10 \\
\ce{H1He1} & 1 & +1 & 6-31G** & 20 \\
\ce{H1Li1} & 1 & 0 & 3-21G & 22 \\
\ce{H1} & 2 & 0 & STO-3G & 2 \\
\ce{H2C1} & 3 & 0 & 3-21G & 26 \\
\ce{H2O2} & 1 & 0 & STO-3G & 24 \\
\ce{H4C2} & 1 & 0 & STO-3G & 28 \\
\ce{He1} & 1 & 0 & STO-3G & 2 \\
\ce{K1} & 2 & 0 & STO-3G & 26 \\
\ce{N1} & 4 & 0 & STO-3G & 10 \\
\ce{Ne1} & 1 & 0 & STO-3G & 10 \\
\ce{O1} & 3 & 0 & STO-3G & 10 \\
\ce{O2} & 1 & 0 & STO-3G & 20 \\
\bottomrule
\end{tabular}
\end{table}

\bibliography{paper}

\end{document}